# Gate-controlled topological conducting channels in bilayer graphene

Jing Li[1], Ke Wang[2,3], Kenton J. McFaul[4], Zachary Zern[1], Yafei. Ren[2,3], Kenji Watanabe[5], Takashi Taniguchi[5], Zhenhua Qiao[2,3*], Jun Zhu[1,6*]

**Affiliations**

[1]Department of Physics, The Pennsylvania State University, University Park, Pennsylvania 16802, USA.

[2]ICQD, Hefei National Laboratory for Physical Sciences at Microscale, and Synergetic Innovation Center of Quantum Information and Quantum Physics, University of Science and Technology of China, Hefei, Anhui 230026, China.

[3]CAS Key Laboratory of Strongly-Coupled Quantum Matter Physics, and Department of Physics, University of Science and Technology of China, Hefei, Anhui 230026, China.

[4]Department of Electrical Engineering, Grove City College, Grove City, Pennsylvania 16127, USA.

[5]National Institute for Material Science, 1-1 Namiki, Tsukuba 305-0044, Japan.

[6]Center for 2-Dimensional and Layered Materials, The Pennsylvania State University, University Park, Pennsylvania 16802, USA.

*Correspondence to: jzhu@phys.psu.edu (J. Zhu) and qiao@ustc.edu.cn (Z. H. Qiao)

**The existence of inequivalent valleys $K$ and $K'$ in the momentum space of two-dimensional hexagonal lattices provides a new electronic degree of freedom, the manipulation of which can potentially lead to new types of electronics, in analogy to the role played by electron spin [1-3]. In materials with broken inversion symmetry, such as an electrically gated bilayer graphene[4,5], the momentum-space Berry curvature $\Omega$ carries opposite sign in the $K$ and $K'$ valleys. A sign reversal of $\Omega$ along an internal boundary of the sheet gives rise to counter-propagating one-dimensional conducting modes encoded with opposite valley indices. These metallic states are topologically protected against backscattering in the absence of valley-mixing scattering, and thus can carry current ballistically[1,6-11]. In bilayer graphene, the reversal of $\Omega$ can occur at the domain wall of AB and BA stacked domains[12-14], or at the line junction of two oppositely gated regions[6]. The latter approach can provide a scalable platform to implement valleytronic operations such as valves and waveguides [9,15], but is technically challenging to realize. Here we fabricate a dual-split-gate structure in bilayer graphene and demonstrate transport evidence of the predicted metallic states. They possess a mean free path of up to a few hundred nanometers in the absence of a magnet field. The application of perpendicular magnetic field suppresses backscattering significantly and enables a 400-nanometer-long junction to exhibit conductance close to the ballistic limit of 4 $e^2$/$h$ at 8 Tesla. Our experiment paves the path to the realization of gate-controlled ballistic valley transport and the development of valleytronic applications in atomically thin materials.**

Exploiting the valley degree of freedom in hexagonal lattices may offer an alternative pathway to achieving low-power-consumption electronics. Experiments have shown that a net valley polarization in the material can be induced by the use of circularly polarized light[2,16,17] or a net bulk current[18-20]. However the use of light is not always desirable in electronics and device proposals using bulk valley polarization often put stringent requirements on the size and edge orientation of the active area[3]. Alternatively, electrically created, valley-polarized topological conducting channels in high-mobility bilayer graphene may offer a robust, scalable platform to realize valleytronic operations[1,6,8-15]. Figure 1a illustrates the dual-split-gating scheme proposed by Martin et al[6], where an AB-stacked bilayer graphene (BLG) sheet is controlled by two pairs of top and bottom gates separated by a line junction. The device operates in the regime where both the left and the right regions of the BLG sheet are insulating due to a bulk band gap induced by the independently applied displacement fields $D_L$ and $D_R$. In the "odd" field configuration, where $D_L D_R < 0$, theory predicts the existence of eight conducting modes (referred to as the "kink" states) propagating along the line of interlayer electrostatic potential difference $V = 0$ (See Supplementary Section 1 for simulations). Each valley supports four chiral modes (two resulting from spin degeneracy and two resulting from layer number) with modes from different valleys counter-propagating as illustrated in Fig. 1a. Their wave functions overlap in real space (see Fig. 1b) but are orthogonal in the absence of short-range disorder. In such cases, backscattering is forbidden and the junction is expected to exhibit a quantized conductance of 4 $e^2/h$. In comparison, in the "even" field configuration, i. e. $D_L D_R > 0$, the junction is expected to be insulating since no kink state is present. This sharp contrast thus enables a clear and convincing demonstration of the existence of the kink states.

The above proposal poses a number of fabrication challenges including the precise alignment of the four split gates shown in Fig. 1a and the necessity of hexagonal boron nitride (h-BN) encapsulation to achieve high sample quality needed to suppress sub-gap conduction due to disorder[5,21] In this work, we have overcome these obstacles and present transport evidence of the kink states. The devices are made by sequentially stacking h-BN, BLG and h-BN atop multi-layer graphene split bottom gates supported on a $SiO_2$/doped Si substrate[22]. The fabrication and alignment procedures are detailed in Methods and Supplementary Section 2. A false-color Scanning Electron Micrograph (SEM) image of a finished device is shown in Fig. 1c, with the junction area highlighted in Fig. 1d. We can align the top and bottom splits to better than 10 nm in general. Results reported here are based on two devices. Their junction widths $w$ and lengths $L$ are respectively $w$ = 70 nm, $L$ = 1 μm for device 1 and $w$ = 110 nm, $L$ = 400 nm for device 2. Measurements of the bulk regions yield high carrier mobilities $\mu$ of 100,000 cm$^2$/Vs and 22,000 cm$^2$/Vs respectively for devices 1 and 2, in comparison with $\mu$ of a few thousand cm$^2$/Vs on oxide-supported samples[5]. The characteristics of the devices and the measurement techniques are described in detail in Supplementary Section 3. Measurements are performed at $T$ = 1.6 K unless otherwise noted.

The high quality of the devices ensures insulating behavior of the bulk BLG when the Fermi level $E_F$ resides inside the bulk band gap $\Delta$, which is a few tens of meV in our experiment. This is evident in Fig. 1e, where we plot the bulk charge neutrality point (CNP) resistance $R_{bulk\ CNP}$ vs the displacement field $D$. It is more than 10 MΩ in the $D$-field range of our measurements, rendering its contribution to the measured junction conductance negligible.

Figure 2 presents the experimental observations of the kink states in device 1. We measure the junction conductance $\sigma_j = 1/R_j$ as a function of the silicon backgate voltage $V_{Si}$, which controls the Fermi level $E_F$ in the junction, at a series of fixed $D_L$ and $D_R$ values of both polarities. As an example, Fig. 2a plots $\sigma_j$ as functions of $D_R$ and $V_{Si}$ at $D_L$ = -0.25 V/nm (upper panel) and +0.25 V/nm (lower panel). The diagonal features connecting the lower left to the upper right corners of the panels correspond to the charge neutrality region of the line junction, whose dependence on $D_R$ and $D_L$ is attributed to a slight misalignment between the top and bottom gates (See Supplementary Section 4). It is immediately clear from the data that $\sigma_j$ near the CNP of the junction is high in the odd field configuration (in white) but low in the even field configuration (in blue). This difference is further illustrated in Fig. 2b, where we plot $\sigma_j$ cut along the two yellow dashed lines drawn in Fig. 2a. $\sigma_j$ is high in both configurations when $E_F$ is outside the bulk band gap. However, when $E_F$ is inside the band gap, $\sigma_j$ decreases to less than 1 µS in the even configuration, but remains high in the range of 10 - 15 µS in the odd configuration, indicating the presence of additional conducting channels. Such disparate behavior of $\sigma_j$ in the even and odd field configurations is systematically observed. To illustrate this contrast, we plot in Fig. 2c the junction resistance $R_j$ at the CNP of the junction only, taking data from many panels similar to that shown in Fig. 2a and spanning all four polarity combinations of $D_L$ and $D_R$. The clear contrast between the even, i.e. (+ +) and (− −), and odd, i.e. (+ −) and (− +) quadrants of the graph strongly supports the existence of the kink states in the odd field configuration, as expected by theory[6].

In general, $R_j$ of the kink states ranges from 40 to 100 kΩ, which corresponds to a mean free path (MFP) of $L_k$ = 70–200 nm using the Landauer-Büttiker formula $R_j = R_0(1 + L/L_k)$ (1), where $R_0 = h/4e^2$ = 6.5 kΩ is the ballistic resistance limit of the 4-fold degenerate kink states and $L$ = 1 µm is the junction length in device 1[24]. Although a small contact resistance (several kΩ) may originate from the electrode/kink interface (see Supplementary Section 5), the large value of $R_j$ indicates significant backscattering, i.e. inter-valley mixing of the kink states. A MFP of $L_k$ < 200 nm is surprisingly short, given that the bulk 2D MFP $L_{2D}$ in our high-quality devices is also a few hundred nm. Intuitively and experimentally[14], one expects $L_k \gg L_{2D}$ since $L_{2D}$ is dominated by small-momentum transfer, intra-valley scattering events caused by long-range Coulomb impurities whereas backscattering between kink states of different valleys requires large momentum transfer, which should occur much less frequently, especially in h-BN encapsulated clean samples[25]. Lattice defects cannot account for our observations; their rare occurrence is confirmed by the high mobility of the BLG bulk and experimentally determined resonant impurity scattering amplitude[26]; the large wave function spread as depicted in Fig. 1b also renders the kink states rather insensitive to scattering by point defects [9].

Although the nature of the backscattering mechanisms remains to be understood, we note several potentially relevant aspects of our experiment. Band structure calculations (See Supplementary Section 7) show that a smoothly varying electrostatic potential profile such as the one we realized (see Fig. 1b) supports, in addition to the kink states which are chiral in each valley, non-chiral states the energies of which protrude into the bulk band gap $\Delta$, as shown in Fig. 3a[9,10]. These states are bound by the width of the junction but delocalized along its length. Their presence effectively reduces the size of the bulk gap $\Delta$, which is only a few tens of meV in our experiment. Coulomb disorder potential, as well as geometrical variations of the lithographically defined junction, can cause additional local variations of the energy of the non-chiral states, resulting in their co-existence with the kink states within a larger energy range than suggested by band structure calculations. At energies close to the CNP of the junction, the non-chiral states can potentially exist in the form of quantum dots. As illustrated in Fig. 3c, the co-existing non-chiral states can potentially enable kink states from different valleys to mix via multiple-particle processes, thus introducing backscattering. Indeed, several such mechanisms have been put forward to explain the imprecise conductance quantization of the quantum spin hall edge states[27] including the possibility of a Kondo effect [28-31]. Similar mechanisms may be of relevance here and will be the subject of future studies. Furthermore, quantum dots formed by non-chiral states can also provide an explanation for the oscillations seen in the $\sigma_j$ traces shown in Fig. 2b, and their *I-V* characteristics and temperature dependence, which we show and discuss in detail in Supplementary Section 6.

We perform numerical studies of the junction conductance using the Landauer-Buttiker formula and the Green's function method (see Supplementary Section 8) with on-site Anderson disorder in the energy range of [-$W/2$, $W/2$], where $W$ measures the disorder strength. Although the Anderson disorder may not reflect all potential valley-mixing mechanisms in experimental samples, it provides an efficient means to model the conductance loss from inter-valley scattering and allows us to examine the effect of the controlling factors of the experiment such as the junction width $w$, the bulk gap size $\Delta$, and the Fermi level $E_F$ of the junction. A few key findings are highlighted in Fig. 3 with the complete results given in Supplementary Section 8. Figure 3a plots the calculated band structure of device 1 with a bulk band gap $\Delta = 30$ meV. $\Delta$ is effectively reduced to ~ 21 meV due to the presence of the non-chiral states. Figure 3b plots three length-dependent junction conductances $\sigma(L)$ corresponding to different Fermi levels $E_1 = 0$, $E_2 = 5$ and $E_3 = 14$ meV (dashed lines indicated in Fig. 3a). Fitting to Eq. (1) yields the MFP of the kink states $L_k$=266, 210, and 141 nm respectively for $E_F = 0$, 5, and 14 meV. We attribute the trend of decreasing conductance $\sigma$ with increasing $E_F$ to the availability of more inter-valley scattering paths that involve the non-chiral states as $E_F$ moves towards, or resonates with the energy of a non-chiral state. This situation is illustrated in Fig. 3c. Our calculations also show that $L_k$ increases with increasing $\Delta$ and decreasing $w$, pointing to effective ways of further improving the ballisticity of the kink states.

The application of a perpendicular magnetic field $B$ effectively suppresses the backscattering process of the kink states. As Fig. 4a shows, the resistance of the kink states $R_{kink}$ decreases rapidly with increasing $B$ in both devices. $R_{kink}$ reaches about 15 kΩ in device 1 at $B \sim 7$ T, suggesting a MFP of approximately $L_k \sim 0.8$ µm. This is a factor of 4-10 increase compared to $L_k$ at $B = 0$ in this device. With a shorter junction length of 0.4 µm, $R_{kink}$ of device 2 reaches and remains in the vicinity of the ballistic limit of $h/4e^2 = 6.5$ kΩ in the field range of 8 to 14 T. In Fig. 4b, we plot $R_j$ ($V_{Si}$) of device 2 in all four displacement field polarities at $B = 8$ T. Near the CNP of the junction, $R_j$ ranges approximately 6 kΩ to 10 kΩ in the (+ –) and (– +) field configurations. $R_j$ is also insensitive to the magnitude of the displacement field, which varies from 0.3 to 0.5 V/nm. In stark contrast, resistances of hundreds to thousands of kΩ are observed in the two even field configurations. These observations are quite remarkable and clearly attest to the presence of nearly ballistic kink states in the expected gating configurations.

The observed magneto-resistance of the kink states is satisfactorily reproduced by our numerical studies, as shown in Figs. 4c and 4d (see Supplementary Section 9). Two effects of the magnetic field may play important roles. As Fig. 4c shows, the formation of Landau levels lifts the non-chiral states away from the energy range of the kink states, which should reduce the probability of any non-chiral state mediated backscattering processes discussed earlier. In addition, the magnetic field exerts Lorentz force of opposite directions on the counter-propagating kink states, causing their wave functions to separate in space. This separation does not occur at $E_F = 0$ but widens with increasing $E_F$ and is more pronounced in a wide junction[10,32]. It also increases with increasing $B$ and can reach a size comparable to the spread of the wave function itself at moderate field strengths (see inset of Fig. 4d). The physical separation of the counter-propagating kink states can strongly suppress any type of valley-mixing mechanisms [28-31], thus resulting in a robust topological protection of the kink states that resembles the chiral edge state picture in the quantum Hall effect.

What we have demonstrated in this work – the creation of one-dimensional topological conducting channels in bilayer graphene using electrical control – opens up exciting new avenues of implementing valley-controlled valves, beam splitters, and waveguides to control electron flow in high-quality atomically thin materials [9,15]. Narrower junctions combined with the use of a larger band gap can potentially enable the kink valleytronic devices to operate at non-cryogenic temperatures. The dual-split-gate structure demonstrated here will enable the realization of potential-controlled few-electron quantum dots[33] and open the door to the exploration of the fascinating edge state and domain wall physics in the quantum Hall regime of bilayer graphene[32,34].

## Acknowledgments

J. L., Z. Z. and J. Z. are supported by ONR under grant No. N00014-11-1-0730 and by NSF grant No. DMR-1506212. K. J. M. is supported by the NSF NNIN REU program in 2013. K. Wang, Y. R. and Z. Q. are supported by the China Government Youth 1000-Plan 342 Talent Program, the Fundamental Research Funds for the Central Universities (Grant Nos.: WK3510000001, WK2030020027), the National Natural Science Foundation of China (Grant No. 11474265), and the National Key R & D Program (Grant No.2016YFA0301700). K. Watanabe and T. T. are supported by the Elemental Strategy Initiative conducted by the MEXT, Japan and a Grant-in-Aid for Scientific Research on Innovative Areas "Science of Atomic Layers" from JSPS. T.T. is also supported by a Grant-in-Aid for Scientific Research on Innovative Areas "Nano Informatics"(Grant No. 25106006) from JSPS. Part of this work was performed at the NHMFL, which was supported by NSF through NSF-DMR-0084173 and the State of Florida. The Supercomputing Center of USTC is gratefully acknowledged for the high-performance computing assistance. The authors acknowledge use of facilities at the PSU site of NSF NNIN. We are grateful for helpful discussions with Rui-rui Du, Herbert Fertig, David Goldhaber-Gordon, William Halperin, Shahal Ilani, Jainendra Jain, Eun-ah Kim, Chaoxing Liu, Xiao Li, Allan H. MacDonald, Qian Niu, Arun Paramekanti and Andrea Young. We thank Jan Jaroszynski of the NHMFL for experimental assistance.

## Author contributions

J. Z. and J. L. conceived the experiment. J. L. designed and fabricated the devices and made the measurements. K. M. assisted in optimizing the procedure used to fabricate the bottom split gates. J. L. and Z. Z. performed the COMSOL simulations. J. L. and J. Z. analyzed the data. K. Wang, Y. R. and Z. Q. did the theoretical calculations. K. Watanabe and T. T. synthesized the h-BN crystals. J. L., J. Z., K. Wang, Y. R. and Z. Q. wrote the manuscript with input from all authors.

## Additional Information

Supplementary information is available in the online version of the paper. Reprints and permission information is available online at www.nature.com/reprints. Correspondence and requests for materials should be addressed to J. Z. and Z. Q.

## Competing financial interests

The authors declare no competing financial interests.

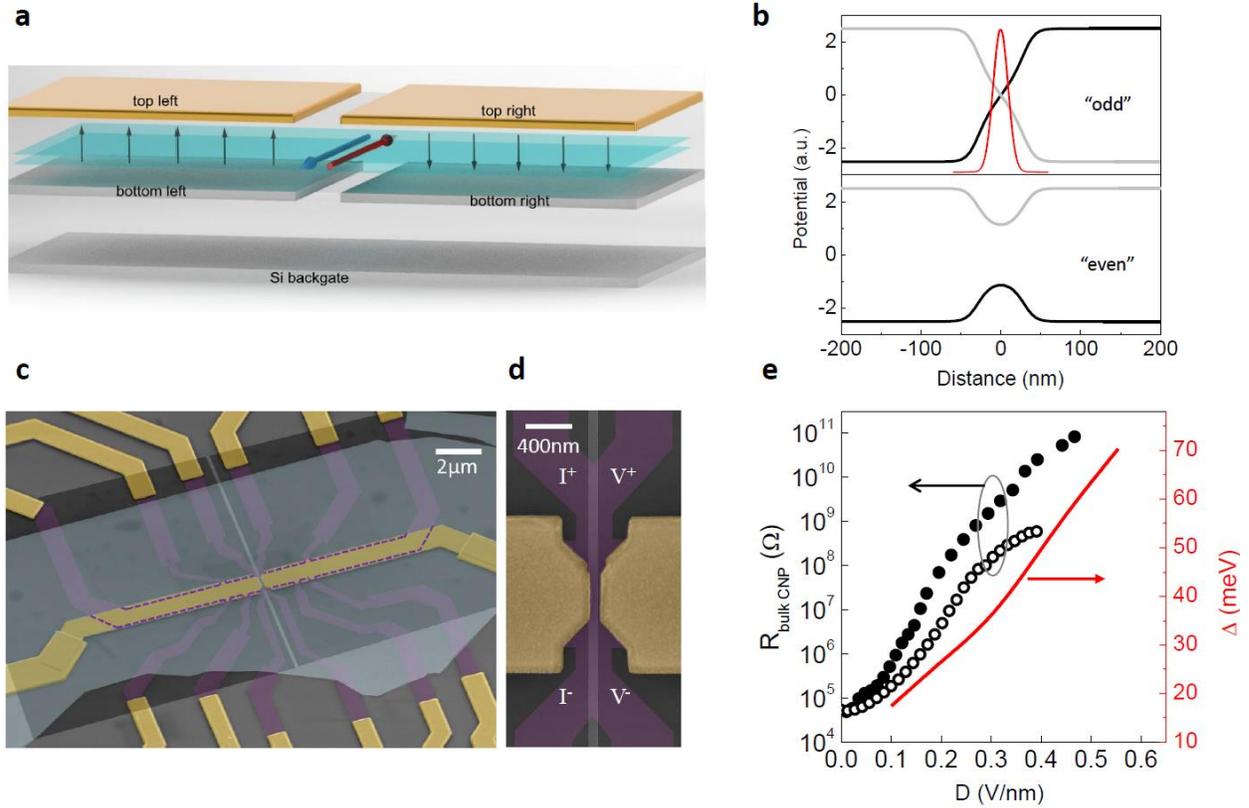

**Figure 1 | Device structure and characterization. a,** Schematic of our dual-split-gated bilayer graphene device. The four split gates independently control the bulk displacement fields $D_L$ and $D_R$ on the left and right sides of the junction. The Si backgate tunes the Fermi energy $E_F$ of the line junction. The gating efficiencies of the split gates are determined using the quantum Hall effect. We determine the gate voltages corresponding to the $D = 0$ and $n = 0$ state on the left and right sides of the junction using the global minima of the charge neutrality point resistance $R_{bulk\ CNP}$. Subsequent measurements are done at $n_L = n_R = 0$ and constant displacement fields $D_L$ and $D_R$. The diagram shows the odd field configuration that results in the presence of the helical kink states at the line junction. Blue and red arrows correspond to modes carrying valley index $K$ and $K'$ respectively. Each one contains four modes accounting for the spin and layer degeneracy. **b,** External electrostatic potential profile near the top (gray) and bottom (black) graphene layers for the odd ($D_L D_R < 0$) and even ($D_L D_R > 0$) field configurations. Potential simulations are performed using the COMSOL package and parameters of device 1. The crossing of the potentials at $V = 0$ gives rise to the topological kink states. The red curve plots the wave function distribution of one such state schematically, with a full width at half maximum of 22 nm. **c,** A false-color SEM image of a device similar to device 2. The bilayer graphene is shaded and outlined in purple, the top gates and electrodes gold, the bottom multi-layer graphene split gates black, and the top h-BN dielectric layer gray. The bottom h-BN layer extends beyond the entire image. **d,** A close-up view of the junction area from another device similar to device 2. The junction is connected to four bilayer graphene electrodes and the measurements use a quasi-four-terminal geometry as shown in the image to eliminate the electrode resistance. Alignment of the gates is generally better than 10 nm. **e,** The bulk charge neutrality point resistance $R_{bulk\ CNP}$ as a function of the applied displacement field $D$ for device 1 (solid symbols) and 2 (open symbols)

in a semi-log plot (left axis). See Supplementary Figure 5 for the device schematic and electrodes used in the measurements. $R_{\text{bulk CNP}}$ rises much more rapidly with the increase of $D$ compared to oxide-supported samples and is larger than 10 MΩ in the range of measurements below. Also plotted on the right is the $D$-dependent bulk band gap $\Delta$ obtained from temperature dependence measurements of a device similar to device 2 (ref. 34).

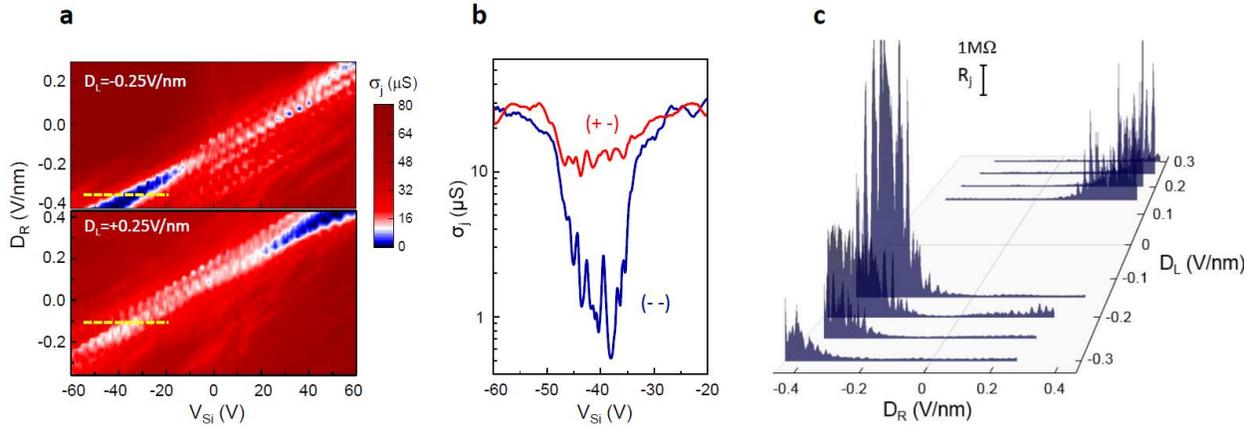

**Figure 2 | Evidences of kink states. a**, The junction conductance $\sigma_j$ as a function of $V_{si}$ at fixed values of $D_R$ from –0.4 V/nm to 0.4 V/nm. Upper panel: $D_L = -0.25$ V/nm. Lower panel: $D_L = +0.25$ V/nm. From device 1. The diagonal bands in the plots correspond to the CNP of the line junction. **b**, $\sigma_j$ vs $V_{Si}$ along the yellow dashed lines marked in the upper (blue curve) and lower (red curve) panels of **a**. We estimate the energy range of the bulk band gap $\Delta$ here corresponds to roughly 25 V on $V_{Si}$. The presence of the kink states in the (+ –) field configuration (red curve) gives rise to high conductance inside the band gap while $\sigma_j$ is low in the (– –) configuration (blue curve). **c**, Junction resistance $R_j$ at the CNP of the junction as a function of $D_L$ and $D_R$ in all four field polarities showing systematically high resistances in the even configurations and low resistances in the odd configurations.

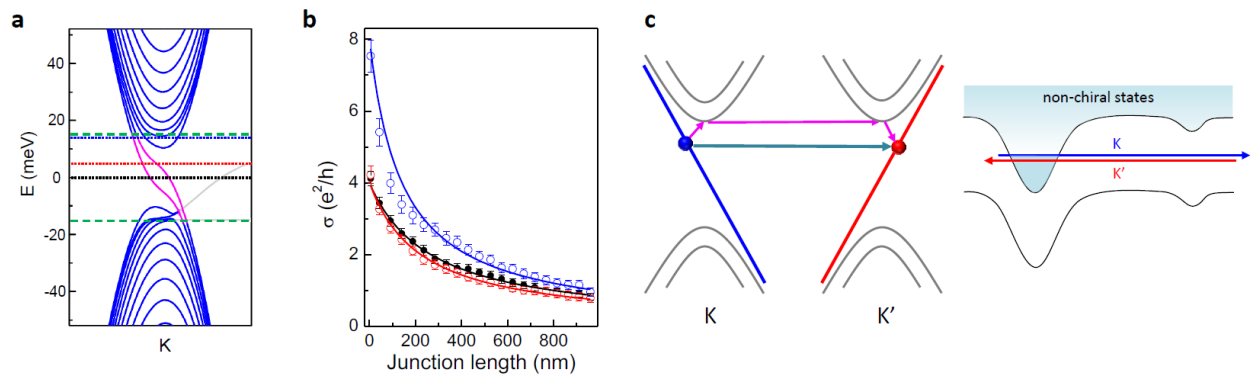

**Figure 3 | Calculated band structure and conductance of the kink states. a,** Band structures of the junction in device 1 ($w = 70$ nm) calculated using COMSOL-simulated potential profiles shown in Fig. 1b. Only the $K$ valley is shown. Non-chiral states bound at the junction reside inside the bulk band gap marked by the green dashed lines. $\Delta = 30$ meV. The kink states are shown in magenta. The gray line corresponds to quantum valley-Hall edge states at the zigzag boundary of the numerical setup, which do not survive edge disorder in realistic samples. **b,** Junction conductance $\sigma$ vs length $L$ calculated at $E_F = 0$ (black), 5 (red), and 14 meV (blue) as marked by the dashed lines in **a**. The disorder strength is chosen to be $W = 0.6$ eV. One non-chiral state is assumed to contribute conductance $4e^2/h$ at $L = 0$. Over 30 samples are averaged for each data point. Error bars are smaller than the symbol size. Fits to Eq. (1) yield MFP of 266, 223, and 141 nm, respectively. The proximity to non-chiral states leads to enhanced backscattering. **c,** An illustration of inter-valley scattering between the kink states of $K$ and $K'$ valleys. A kink state may be directly scattered to a different valley or scattered via coupling to non-chiral states. Non-chiral states can also form quantum dots due to Coulomb potential fluctuations and co-exist with the kink states over a large energy range, as shown schematically.

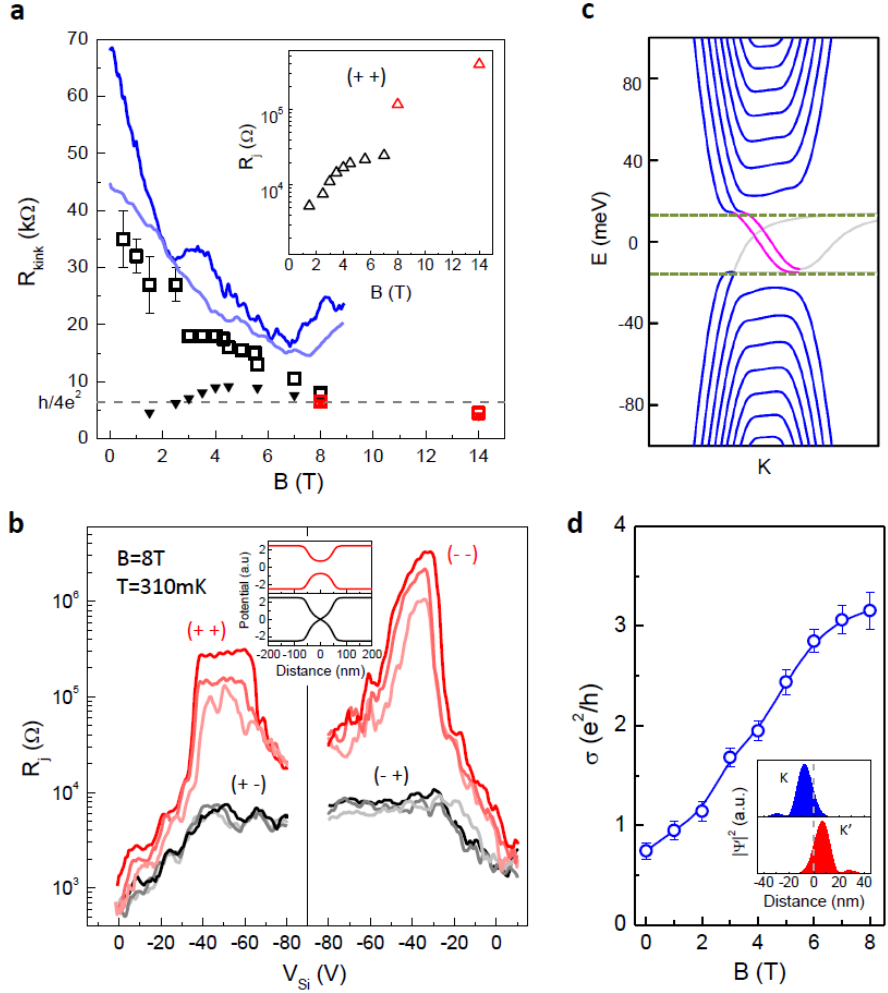

**Figure 4 | Kink state resistance in a magnetic field. a,** Two representative magnetoresistance traces from device 1. $D_L = +0.2$ V/nm and $D_R = -0.3$ V/nm, $V_{Si} = -55.5$ V and -39.3 V for the blue and light blue curve respectively). See Supplementary Section 10 for the details of the measurements. Symbols are from Device 2. The solid triangles are raw magnetoresistance data of $R_j$ in the (+ -) field configuration. $R_j$ can be smaller than $h/4e^2$ due to parallel conduction of the non-chiral states. We use a two-channel model to estimate the resistance of the kink states, using $R_j$ of the (+ +) field configuration shown in the inset to approximate the resistance of the parallel channel. The open squares in **a** plot the estimated resistance of the kink states. At large magnetic field, the non-chiral states become sufficiently insulating that the raw $R_j$ measures directly the kink state resistance. Band structure calculations and detailed discussions regarding the two-channel model are given in Supplementary Section 10. Black symbols: $T = 1.6$ K. Red symbols: $T = 310$ mK in a separate cool-down. $|D_L| = |D_R| = 0.3$ V/nm. **b,** $R_j$ versus $V_{Si}$ in device 2 at $B = 8$ T for all four field configurations as marked in the plot. From dark to light colors: $|D_L| = |D_R| = 0.5, 0.4$ and $0.3$ V/nm. Inset: Potential profile for even (red) and odd (black) field configurations. **c,** The band structure of device 1 shown in Fig. 3a recalculated at $B = 6$ T. The olive dashed lines mark the edges of the bulk conduction and valance bands in Fig. 3a. Non-chiral states residing below the band edges are now lifted to higher energies. The kink states are shown in magenta and the zigzag edge states are in gray (not relevant in realistic samples). **d,** The calculated

magneto-conductance for device 1. See Fig. 3b for parameters used in the calculation. $E_F = 5$ meV. Inset: Wave functions of the $K$ (blue) and $K'$ (red) valley kink states at $B = 6$ T showing a spatial separation of 14 nm due to the Lorentz force. The wave function separation is zero at $E_F = 0$ and increases with increasing $E_F$. It also increases with increasing $B$. See supplementary Section 8 for further details.

**Methods**

**Device fabrication.** The fabrication of the structure shown in Fig. 1c involves six steps. In step 1, we fabricate the bottom split gates made of multi-layer graphene exfoliated from Kish graphite. Thin flakes (~2 nm) are exfoliated to $SiO_2$/doped Si substrates with 290 nm of thermal oxide. We use electron beam lithography (EBL) and oxygen plasma etching (Plasma-Therm Versalock oxygen plasma 14 Watt power for ~30 seconds) to pattern the bottom split gate. Resist ZEP 520a (300 μC/cm$^2$ dose, developed in n-amyl acetate, MIBK: IPA 8:1, and IPA at ~4 °C for 30 seconds each) is used as it provides better resolution than PMMA. An example of a finished bottom split gate is shown in Supplementary Figs 3 (a) and 3 (b). The sample is then annealed in Ar/H$_2$ and/or treated in a very gentle oxygen plasma (MetroLine M4L) to remove ZEP residue. In step 2, thin flakes of h-BN (15–30 nm) and bilayer graphene are sequentially transferred to the bottom split gates using a PMMA/PVA stamp[22]. In step 3, the bilayer graphene flake is shaped into a Hall bar with leads using standard EBL and oxygen plasma etching (see Fig. 1c). In step 4, another thin flake of h-BN is transferred atop the stack to cover the Hall bar but not the graphene leads entirely. In step 5, standard EBL and metal deposition (5 nm Ti / 70 nm Au) is used to make electrical contacts to the graphene leads. The resist is slightly over-developed to ensure good ohmic contacts. In step 6, we pattern the top split gates using EBL and metal deposition (5 nm Ti / 20 nm Au). We measure the width of each pair of bottom split gates using SEM after ZEP removal. To obtain top split gates of matching width, we use empirical relations between the designed width and the actual width after exposure and lift-off, e.g. a split designed to be 100 nm comes out to be around 70 nm due to evaporation angle and the proximity effect of the e-beam dose. Developing the top split gate pattern in ice bathed developer at ~4 °C (with a higher e-beam dose of 450 μC/cm$^2$) provides better control of the development process and hence the dimension of the top split. Split widths $w$ down to 50 nm can be made this way, with higher yield for $w > 70$ nm due to occasional shorting along the splits. An example of the top split gates is shown in Supplementary Fig. 3 (c). To ensure the high quality of the sample, we anneal the h-BN top surfaces in Ar/O$_2$ (90/10%, 500 sccm) at 450 °C for 3 hours and the graphene surfaces in Ar/H$_2$ (90/10%, 500 sccm) at 450 °C for 3 hours before each transfer to remove the polymer residue from previous transfer or lithography[35]. The layer-by-layer transfer approach ensures that both the top and bottom gates extend beyond the bilayer sheet, creating a uniform density profile and ensuring the bulk of the bilayer becomes very insulating in the gapped regime. We avoid aligning the line junction with a straight flake edge that may suggest either the zigzag or the armchair orientation. The alignment of the top and bottom splits is critical to the success of the experiment. While relying on the same pre-patterned alignment markers (made by GCA 8000 stepper in our case, squares with 20 μm on the side) suffices for most multi-step EBL fabrications that require alignment, the misalignment between the two pairs of split gates can be up to 90 nm in random directions even when an advanced e-beam writer with precise stage movement such as ours (Vistec EBPG5200) is used. The error primarily comes from the imperfection of the alignment markers made by optical lithography. We have employed a realignment procedure to address this issue. In step 1 of the lithography, dummy graphene splits aligned with the bottom split gates were made (boxed area in Supplementary Fig. 4 (a). In step 5 when we pattern the Ti/Au electrodes, we also pattern metal splits designed to overlap with the dummy graphene splits as shown in Supplementary Fig. 4 (b). In the same step, we write a second set of alignment markers that is designed to track the center of the bottom split gates. As Supplementary Fig. 4 (c) shows, the metal splits are slightly shifted from the dummy graphene splits (20 nm to the right in this case but can be up to 90 nm). In step 6 where we pattern the top split gates, we use the alignment markers made by EBL in step 5, which has more precise dimensions and make corrections for the misalignment determined in Supplementary Fig. 4 (c). As a final check, we make another pair of dummy metal splits concurrently with the top split gates in step 6. This is shown in Supplementary Fig. 4 (d). It is indeed aligned with the dummy graphene split underneath. Using this procedure, we can align the top and bottom split gates to better than 10 nm reliably. The width of the top and bottom split gates can be matched to better than 15 nm.

# Supplementary Information for

# Gate controlled topological conducting channels in bilayer graphene

Jing Li, Ke Wang, Kenton J. McFaul, Zachary Zern, Yafei. Ren, Kenji Watanabe,
Takashi Taniguchi, Zhenhua Qiao[*], Jun Zhu[*]

*Correspondence author.
E-mail: jzhu@phys.psu.edu (J. Zhu) and qiao@ustc.edu.cn (Z. H. Qiao)

Supplementary Text Contents

1. Electrostatics simulations using COMSOL Multiphysics[®]

2. Device fabrication

3. Device characterizations

4. Tracking the charge neutrality point (CNP) of the line junction in device 1

5. Schemes of contacting the kink states

6. Oscillations, *I-V* characteristics and the temperature dependence of the junction conductance

7. Band structure calculations of the biased bilayer graphene junction

8. Calculating the conductance of the kink states

9. Modeling the effect of a perpendicular magnetic field

10. Determining the conductance of the kink states in a perpendicular magnetic field

# 1. Electrostatics simulations using COMSOL Multiphysics®

COMSOL Multiphysics® is a software platform which utilizes finite element analysis to simulate various physics-based problems. We use COMSOL Multiphysics® (version 4.4) stationary electrostatics package to simulate the external electrostatic potential profile near the line junction and the gating efficiency of the Si backgate.

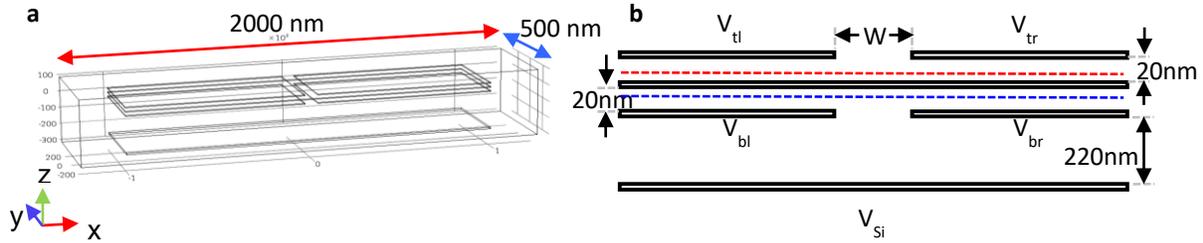

**Supplementary Figure 1** | (a) Schematics of device structure constructed in COMSOL. (b) Side view of the five gates and the bilayer graphene sheet near the splits with dimensions marked in the figure (not drawn to scale). The external electrostatic potential $U$ computed along the red and blue dashed lines is shown in Supplementary Fig. 2.

Supplementary Fig. 1 (a) and (b) show the setup in COMSOL and the dimensions of the structure simulated. Split width $w$ = 70 nm (device 1) or 110 nm (device 2). All five gates and the bilayer graphene are represented by 5 nm thick metal plates. The whole device is enclosed in a dielectric environment of $\varepsilon$ = 3 that describes our h-BN flakes. The SiO$_2$ thickness is set to be 220 nm to account for the smaller $\varepsilon$ used (actual thickness 290 nm and $\varepsilon \sim$ 3.9).

Supplementary Fig. 2 (a) - (d) plot the external electrostatic potential $U_{top}$ along a line cut of 0.1 nm above the bilayer (red dashed line in Supplementary Fig. 1 (b)) and $U_{bot}$ along a line cut of 0.1 nm below the bilayer (blue dashed line in Supplementary Fig. 1 (b)) in devices 1 and 2 for both even and odd bias configurations. Without considering the screening of the bilayer explicitly, the spatial dependence of $U_{top}$ and $U_{bot}$ approximates that of the top and bottom layer graphene potential respectively across the junction. The bulk band gap $\Delta$ is approximately $U_{top}$-$U_{bot}$. In the even configuration, both $U_{top}$ and $U_{bot}$ approach zero in the junction; the gap shrinks but remains finite. As the comparison between devices 1 and 2 shows, a narrower split (device 1) can maintain a large gap in the junction. In the odd configuration, $U_{top}$ and $U_{bot}$ intersect at the zero potential line (through the paper), where the kink states are predicted to exist. The finite width of the junction causes $U_{\text{top}}$ and $U_{\text{bot}}$ to change smoothly across the junction. As our calculations show (Supplementary Fig. 2), the smooth potential profile gives rise to additional non-chiral states bound at the junction, the energy of which extends into the bulk band gap. The energy proximity of the non-chiral states plays an importance role in the backscattering of the kink states.

We have also simulated the gating efficiency of the silicon backgate $\alpha_{\text{Si}}$ on the junction by simulating the $V_{\text{Si}}$ dependent carrier density in the junction while the split gates are fixed at voltage described in the caption of Supplementary Fig. 2. Near the center of the junction, it is found to be $3.5 \times 10^{10}$ cm$^{-2}$V$^{-1}$ in device 1 and $4.8 \times 10^{10}$ cm$^{-2}$V$^{-1}$ in device 2. Both are independent of the voltages applied to the four split gates as expected from electrostatics. Both are smaller than the planar capacitance value of $6.9 \times 10^{10}$ cm$^{-2}$V$^{-1}$, consistent with the junction geometry. Using $\alpha_{Si}$ and a density of states value of $2.5 \times 10^{13}\,\text{cm}^{-2}/\text{eV}$ for bilayer graphene, we estimate that a range of 25 V on the silicon gate corresponds to an energy range of ~ 35 meV, in good agreement with the value of the band gap $\Delta$ at the $D$ field strength used in Fig. 2 of the main text.

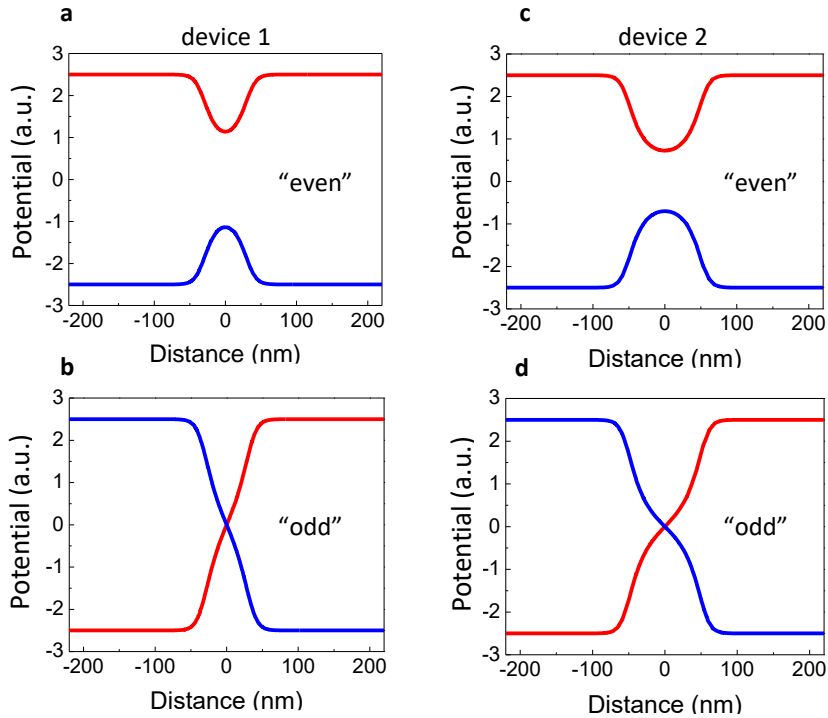

**Supplementary Figure 2** | $U_{\text{top}}$ (red) and $U_{\text{bot}}$ (blue) along the line cuts shown in Supplementary Fig. 1 (b) for device 1 ((a) and (b)) and device 2 ((c) and (d)). $V_{\text{tl}} = V_{\text{tr}} = 5$ V, $V_{\text{bl}} = V_{\text{br}} = -5$ V for the even configurations. $V_{\text{bl}} = V_{\text{tr}} = 5$ V, $V_{\text{tl}} = V_{\text{br}} = -5$ V for the odd configurations. $V_{\text{Si}} = 0$ V. $V_{\text{graphene}} = 0$ V.

## 2. Device fabrication

Text description is in Methods.

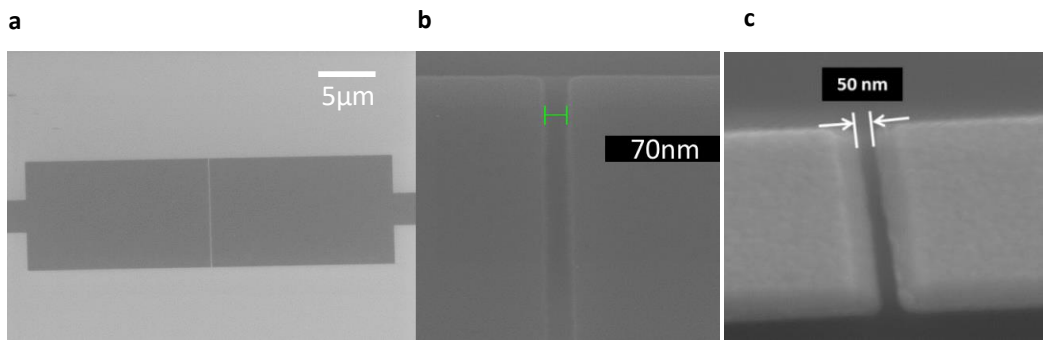

**Supplementary Figure 3** | (a) SEM image of a pair of bottom split gates made of multi-layer graphene. (b) Close-up view of the split in (a). (c) SEM image of a pair of Au top split gates with a split size of 50 nm.

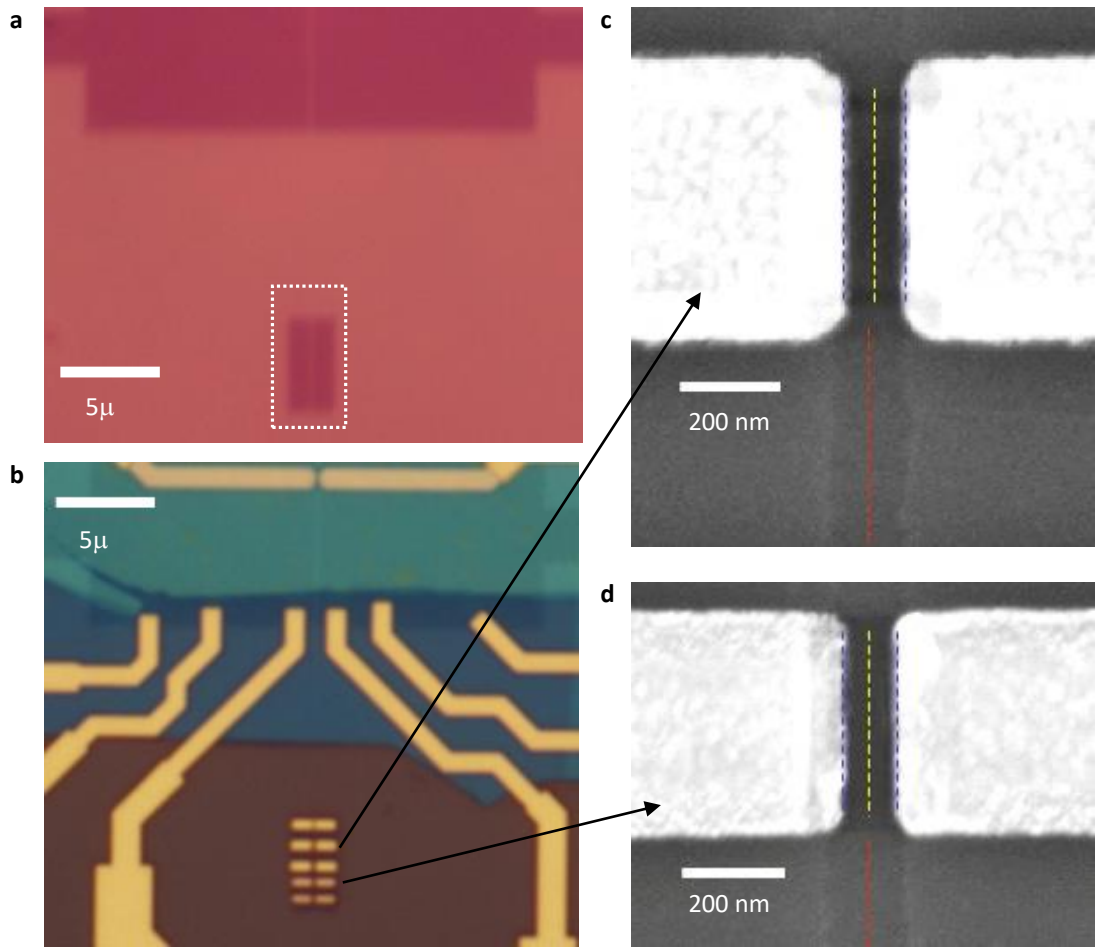

**Supplementary Figure 4** | (a) Optical micrograph of a pair of bottom split gates together with the dummy graphene split (boxed area) aligned to its center. (b) Optical micrograph of the same area after the fabrication is completed showing the metal splits patterned in step 5 (top three pairs) and step 6 (bottom two pairs) on top of the dummy graphene split (c) SEM micrograph of one metal split patterned in step 5. Its center (yellow dashed line) is shifted to the right of the center of the dummy graphene split (red dashed line) by 20 nm. (d) SEM micrograph of a metal split patterned in step 6 showing precise alignment to the dummy graphene split.

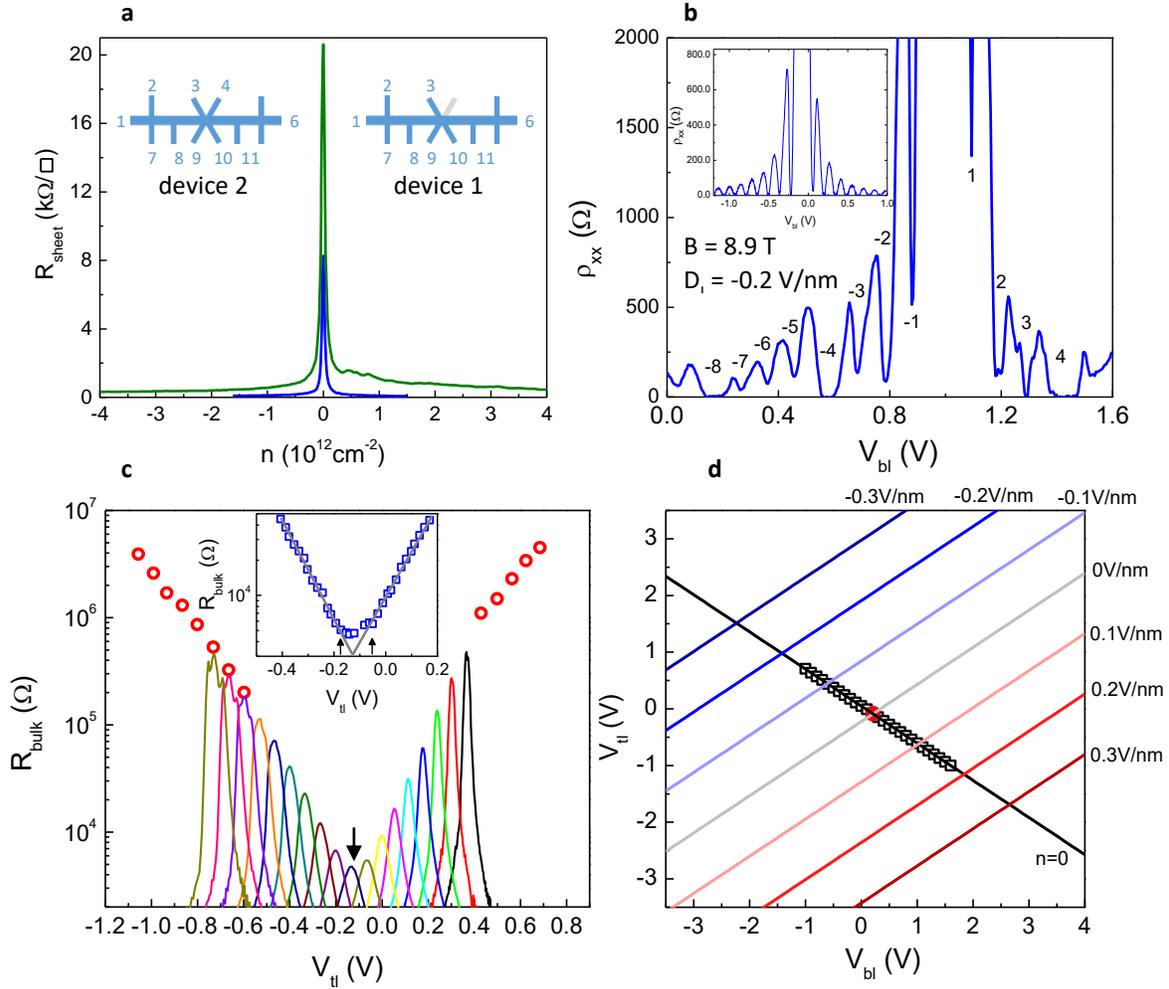

**Supplementary Figure 5** | (a) The bulk sheet resistance vs carrier density for device 1 (blue) and 2 (olive). The gating efficiencies are determined from the quantum Hall effect. The inset shows the schematics of both devices. Measurements are done on the left side of the devices by grounding the bottom gate $V_{bl}$ and sweeping the top gate $V_{tl}$. $V_{tr} = V_{br} = 0$. The sheet resistance is calculated from $R_{16, 78}$ (current from pin 1 to pin 6 and voltage probes 7 - 8) in both cases. (b) $R_{16, 78}$ vs $V_{bl}$ in device 1 showing fully resolved integer quantum Hall states at $B = 8.9$ T. $V_{tl}$ / $V_{bl}$ is swept simultaneously to maintain $D = -0.2$ V/nm. $V_{tr} = V_{br} = $ 2V. Inset: similar measurement at $B = 3$ T, and $D = 0$ V/nm showing the 8-fold degenerate $N = 0$ and 1 Landau levels of bilayer graphene. (c) $R_{16, 78}$ vs $V_{tl}$ at selected $V_{bl}$ from 1.2 V (leftmost curve) to -0.5 V (rightmost curve) decreasing in 0.1 V step. Open circles are two-terminal CNP resistance $R_{78, 78}$ (contact resistances are negligible compared to bulk resistance) at $V_{bl} = 1.7$ V to 1.0 V (left) and at $V_{bl} = -0.6$ to -1.0 V (right). Also in 0.1 V step. The arrow marks the global minimum. The inset shows the CNP resistance near the global minimum. (d) The $n = 0$ line on the $V_{tl}$-$V_{bl}$ plot obtained from the CNP positions in (c). The global minimum of the CNP resistance in (c) corresponds to $n = 0$ and $D = 0$. Constant $D$ lines are indicated in the graph. Data in (b) - (d) are from device 1.

## 3. Device characterizations

Supplementary Fig. 5 (a) plots the sheet resistance vs. carrier density in the bilayer bulk for both devices. The field effect mobility $\mu_{FE}$ is 100,000 cm$^2$V$^{-1}$s$^{-1}$ and 22,000 cm$^2$V$^{-1}$s$^{-1}$ respectively for device 1 and 2. The high quality of the devices is further seen in Supplementary Fig. 5 (b), where the integer quantum Hall states of device 1 are fully resolved at $B$ = 8.9 T and $D$ = -0.2 V/nm. We use the quantum Hall effect to determine the gating efficiency of the top and bottom gates accurately, the results of which are given in Supplementary Table I. The schematics of the devices are given in the inset of Fig. 5 (a). Figure S5 (c) plots the left bulk resistance $R_{bulk}$ ($R_{16,78}$) vs $V_{tl}$ at various fixed $V_{bl}$ in device 1. The CNP resistance starts to rise approximately exponentially with increasing $D$ at a small onset field of $D_{on}$ = 0.01 V/nm (0.07 V/nm for device 2), suggesting small Coulomb potential fluctuations of a few meV caused by electron-hole puddles[1]. Tracking the location of the CNP on the $V_{tl}$-$V_{bl}$ plane allows us to determine the $n$ = 0 line, the $D$ = 0, $n$ = 0 point and constant $D$ field lines, following the definite of $D$ in Ref.[1]. These are shown in Supplementary Fig. 5 (d). We sweep gates simultaneously to follow constant $D$ field lines while keeping the bulk BLG at the CNP. The line junction resistance is measured using a quasi-four-terminal geometry for device 2 ($R_{3,9,410}$) and a quasi-three-terminal geometry for device 1 ($R_{39,310}$) (see inset of Supplementary Fig. 5(a)). The resistance of electrode 3 is estimated to be 3kΩ and subtracted from the three-terminal data. The resulting junction resistance is plotted in Figs. 2 and 4. See Supplementary Section 10 for more discussions of this subtraction and its validity in a magnetic field. We use standard low current excitation lock-in techniques in the low impedance regime (< 1 MΩ). In the high impedance regime, we source a constant voltage and measure the current in a two-terminal geometry. $T$ = 1.6 K unless otherwise noted.

## 4. Tracking the charge neutrality point (CNP) of the line junction in device 1

A combination of slight misalignment and mismatch in the location and width of the top and bottom split gates as illustrated in Supplementary Fig. 6 (a) can cause net doping of the junction area as a function of changing $D_L$ and $D_R$ even when the bulk BLG on both sides is kept at charge neutrality. Consequently the CNP of the junction appears at different $V_{Si}$ for different $D_L$/$D_R$, giving rise to a slope on the $D_R$-$V_{Si}$ plane (See $\sigma_j$ plots shown in Fig. 2a).

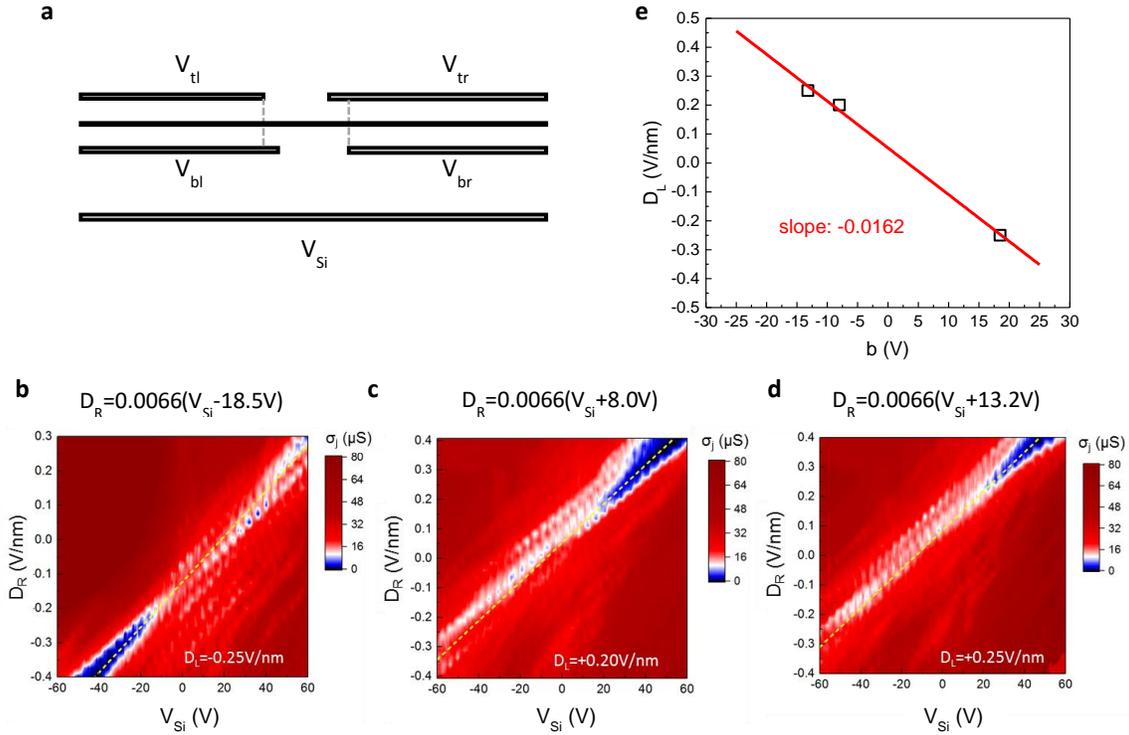

**Supplementary Figure 6** | (a) A illustration of the gate alignment situation in device 1. (b) - (d) Junction conductance $\sigma_j$ as a function of $V_{Si}$ and $D_R$ at fixed $D_L$'s as labeled in the plots. The yellow dashed lines are fits that track the CNP of the junction. (e) $D_L$ vs $b$ using values obtained in (b) - (d).

Supplementary Fig. 6 (b) to (d) show three more $\sigma_j$ plots similar to that of Fig. 2a, with $D_L$ fixed at values from -0.25 V/nm to +0.25 V/nm. , We draw three lines of the same slope through the tip of the blue-colored region (i.e. resistance peak at low $D_R$ field in $V_{Si}$ sweep) and require the lines to pass through the region of the highest resistances at large $D_R$ field (this region corresponds to the black color in the figure but typically contains more than one peak) in each plot. We manually assess the quality of the fits and adjust the slope until a value that works satisfactorily for all three plots is obtained. This process yields three linear fits $D_R = k(V_{Si} + b(D_L))$ with the same slope $k = (+0.0066 \pm 0.0002)$ /nm that is robust from cool down to cool down. The parameter $k$ represents the gating effect of the right gates on the junction. A positive value indicates the top right gate extends into the junction as illustrated in Supplementary Fig. 6 (a). The parameter $b$ represents the gating effect of the left gates, the change of which with $D_L$ yields a slope of -0.0162, indicating the bottom left gate extends into the junction, but not as much as the top right gate. This means the top split is slightly narrower than the bottom split (70 nm) in device 1. We note that in device 2, the kink regime appears at roughly the same range of $V_{Si}$ in spite of the change of $D_L$ and $D_R$, suggesting nearly perfect match of the split gates.

## 5. Schemes of contacting the kink states

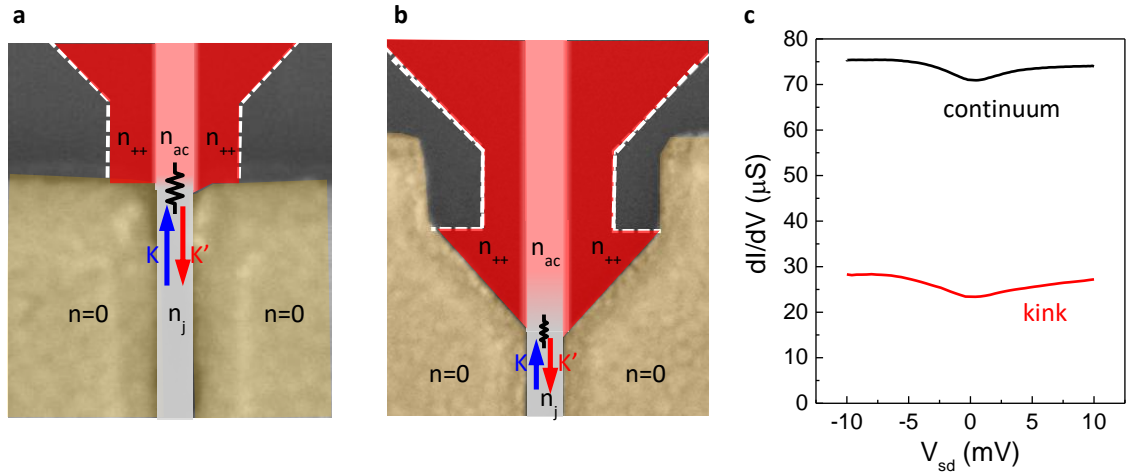

**Supplementary Figure 7** | (a) and (b) Contacting schemes used in device 1 (a) and 2 (b). Regions of different doping are colored on top of SEM images of similar devices. (c) $dI/dV$ vs $V_{sd}$ at an $E_F$ positioned in the heavily hole doped continuum (black trace) and an $E_F$ positioned in the kink regime (red trace, $D_L$ = +0.2 V/nm, $D_R$ = -0.2 V/nm and $V_{Si}$ = -52 V).

Supplementary Fig. 7 (a) and (b) show the two different schemes used to contact the kink states in device 1 (Supplementary Fig. 7 (a)) and device 2 (Supplementary Fig. 7 (b)) respectively, overlaid on top of SEM images of devices of each type. Two leads are patterned at each end of the junction, forming a Y-shaped connection. The leads are made of BLG and are contiguous with the bulk BLG and the junction itself. However, region 1 (colored red) is heavily doped to $n_{++} \sim 2 \times 10^{12}$ cm$^{-2}$ because it is only gated by the back gates and the junction itself is at low doping $n_j$. In between the two an access region exists, where $n_{ac}$ is in between $n_{++}$ and $n_j$. In device 1, the top split gates have close to 90° degree bent at the end of the junction and the change from $n_{ac}$ to $n_j$ is relatively abrupt. We suspect such sharp transition may have led to a junction and consequently a small interfacial resistance $R_{ac}$ that is on the order of a few kΩ. Supplementary Fig. 7 (c) plots the differential conductance $dI/dV$ of a highly conductive kink state (~ 50 k$\Omega$) together with a heavily hole-doped continuum regime. Both show slight non-linear *I-V*. The conductance change of the kink state corresponds to ~ 7 k$\Omega$. This may have been caused by the interface. This resistance was not subtracted from any data presented. Inspired by practices in studying quantum point contacts, the top split gates in device 2 tapers gradually into the junction at a 45° angle to create a smoother transition between $n_{ac}$ and $n_j$ and a "funnel" effect.

# 6. Oscillations, *I-V* characteristics and the temperature dependence of the junction conductance

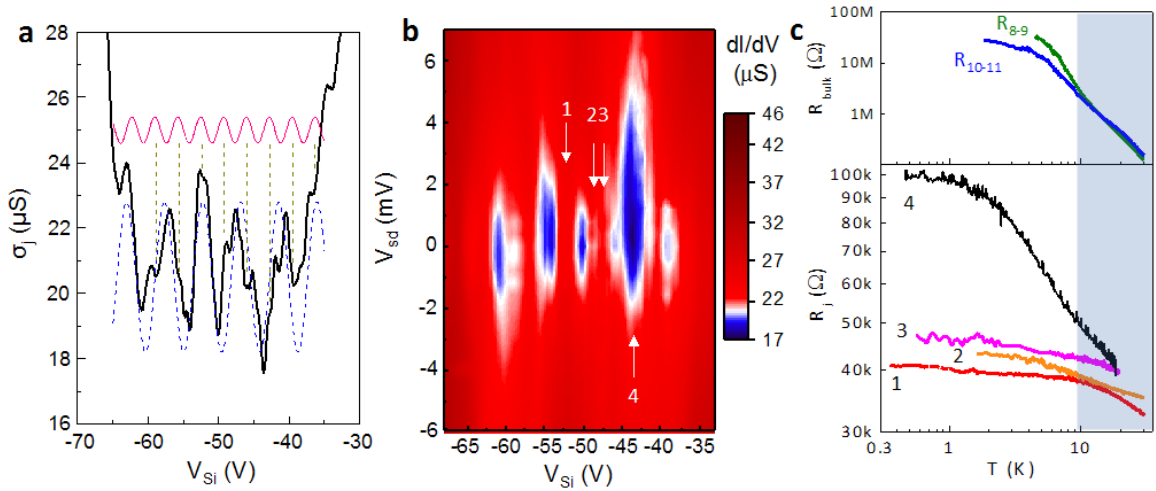

**Supplementary Figure 8** (a) $\sigma_j$ (solid black curve) vs $V_{Si}$. $D_L$ = + 0.20 V/nm and $D_R$ = - 0.20 V/nm. Blue and pink lines are guide to the eye with periods of 5.4 V and 3.2 V respectively. (b) Differential conductance map *dI/dV* of the same regime as in (a). (c) Temperature-dependent junction resistance $R_j$ at $V_{Si}$ marked in (b). Also plotted is the *T*-dependent resistance of the gapped bulk bilayer at the CNP. From Device 1.

A general feature of the kink regime is the appearance of conductance oscillations on the order of a few µS superimposed onto a smooth $\sigma_j$ background. One example is shown in Supplementary Figure 8 (a). Such oscillations are also present in the even field configurations (see Fig. 2 (b) of the text). The oscillations typically consist of two or three prominent periods in $V_{Si}$, as illustrated in Supplementary Fig. 8 (a). Differential conductance map *dI/dV* shown in Supplementary Fig. 8 (b) reveals non-linear *I-V* at low-conductance points with energy scales of a few meV. At the first glance, they resemble coherent conductance oscillations of one-dimensional states in carbon nanotubes with semi-transparent contacts[2], however the details of the oscillations, such as their magnitude and period, vary among different kink regimes and sometimes between cool-downs for the same regime. This variation makes it unlikely that the oscillations originate from the confinement of the kink states themselves. Instead, they can be explained naturally by considering the parallel conduction of localized charge puddles, i.e. quantum dots of the non-chiral states formed due to Coulomb potential fluctuations. In Supplementary Figure 8 (c), we plot the temperature dependence of the junction resistance $R_j$ at selected $V_{Si}$'s as marked in Supplementary Figure 8 (b). States with more non-linear $IV$ characteristics also exhibit stronger temperature dependence. At $T < 10$ K, the T-dependence of the nearly linear kink states is very weak. Above 10 K, the T-dependence of the kink states is difficult to assess due to the onset of parallel conduction through the gapped bulk. More studies are required to understand the role of electron-electron interaction[3] and other predicted temperature dependences[4-6].

## 7. Band structure calculations of the biased bilayer graphene junction

### 7.1. Tight-binding model of biased AB-stacked bilayer graphene with disorder

We model the biased AB-stacked bilayer graphene used in experiment with a π-orbital tight-binding Hamiltonian including a potential difference $2U$ between the top ($U_{top}$) and bottom layers ($U_{bot}$) and Anderson disorder [7]:

$$H_0 = -t \sum_{\langle ij \rangle} \left( c_i^\dagger c_j + \text{H.c.} \right) + t_\perp \sum_{\langle i \in T, j \in B \rangle} \left( c_i^\dagger c_j + \text{H.c.} \right)$$
$$+ \sum_{i \in T} U_{top} c_i^\dagger c_i - \sum_{i \in B} U_{bot} c_i^\dagger c_i + \sum_i \epsilon_i c_i^\dagger c_i,$$
(Eq. S1)

where $c_i^\dagger (c_i)$ is the creation (annihilation) operator of an electron on site $i$. The first and second terms represent the intra- and inter-layer nearest-neighbor hopping, respectively. Hopping energies of $t = 2.6$ eV and $t_\perp = 0.34$ eV are used in our calculations. The third and fourth terms indicate the site energies at the top and bottom layers separately. The last term represents the on-site Anderson disorder with $\epsilon_i$ being randomly distributed in the energy interval of $[-W/2, W/2]$, where $W$ measures the disorder strength.

### 7.2. The band structure of the junction

Supplementary Fig. 9 schematically displays the setup of the bilayer graphene line junction used in our calculations. The total system has a finite width $d$ along the $x$ direction and translational symmetry along the $y$ direction. The junction width is $w$. The interlayer potential difference $U(x)$ is set to be $+2U$ ($-2U$) on the left (right) side of the junction and varies across the junction. In our setup, we use the experimentally determined junction width of both devices (70 nm and 110 nm for device 1 and 2, respectively), while the total system width ($d = 120$ nm) is much narrower than the real device parameters due to the limitation of the computational capacity, but sufficient to capture the main physics of the junction region. Previous studies[8] show that the conductance of the kink states is insensitive to their orientations with respect to crystallographic directions of the graphene lattice. In our calculations, we have examined both the zigzag and the armchair edge orientations of the junction. We find the main results of the calculations, namely, the conductance decay of the kink states due to inter-valley scattering caused by the on-site Anderson disorder, and the magnetic field-induced separation of the wave functions of the kink states in the $K$ and $K'$ valleys, to be insensitive to the edge orientation. Results using the zigzag edge orientation are presented below. The experimental orientation of the kink states is likely to vary along the length of the junction due to size/shape variance of the lithographically defined gates.

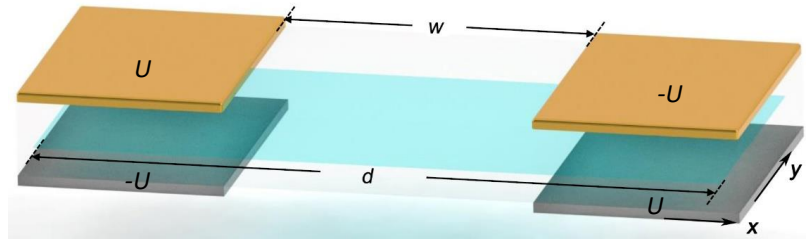

**Supplementary Figure 9** | Schematic of a bilayer graphene line junction in our calculation. The layer potential of the bilayer graphene in the biased regions is set to be $U$ and $-U$ respectively as indicated in the graph. $w$ and $d$ are respectively the junction and the total ribbon widths. $d = 120$ nm in our calculations.

Band structures of the line junction are obtained by exactly diagonalizing the tight-binding Hamiltonian given in Eq. (S1). Results corresponding to different junction widths $w$ from 0 to 60 nm are plotted in Supplementary Fig. 10. For clarity, only the bands near the $K$ point are shown (bands near $K'$ are symmetric to those near $K$). The bulk band gap at the region with constant-$U$ is $\Delta \sim 2U = 100$ meV as marked by green dashed lines in the figure. The opposite-polarity configuration (+2$U$ on the left and -2$U$ on the right) gives rise to the chiral kink states inside the band gap, shown in red. In addition, increasing junction width $w$ leads to the appearance of more states inside the bulk band gap. These states are non-chiral and bound at the junction. Similar results were reported in Ref.[9]. As the Fermi level of the junction $E_F$ moves through the gap, the kink states may co-exist with the non-chiral states at certain energies. Calculations also produced edge modes resulting from the zigzag orientation of the junction as shown in cyan in Supplementary Fig. 10. These zigzag edge states can contribute to a conductance of 4 $e^2/h$ in the absence of disorder. However, their contribution is negligible in experiment due to the random atomic orientations at the junction boundaries. This reality is reflected in our calculation by artificially introducing strong on-site disorder ($W$ = 10 eV) along a ten-atom-wide strip at the boundaries of the junction to eliminate their contribution to the conductance.

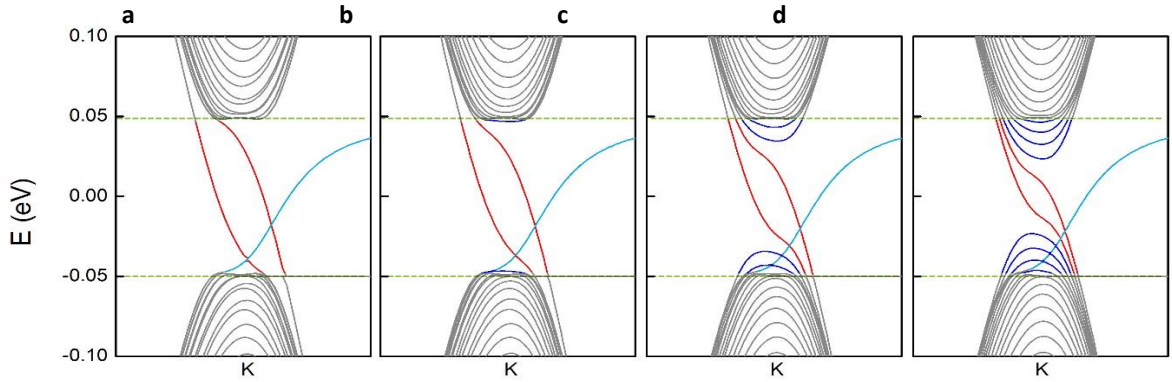

**Supplementary Figure 10** | Band structures of the bilayer graphene line junction calculated with different junction width $w$ = 0 (a), 20 (b), 40 (c), and 60 nm (d). A cosinusoidal potential profile (Supplementary Fig. 10) with $2U$ = 100 meV is used in all four calculations. Red, blue and cyan colors represent the kink, bound and zigzag edge states respectively. The green dashed lines mark the energy range of the bulk band gap. As the junction width $w$ increases, more non-chiral states appear inside the bulk gap. Only $K$ valley is shown here.

**7.3. The influence of the kink potential profile on the band structure**

Supplementary Fig. 11 compares the band structures obtained using three different potential profiles $U(x)$, namely a cosinusoidal, a linear, and the COMSOL-simulated profile for both devices 1 and 2. The interlayer potential difference is set to be $2U$ = 30 meV to mimic the experimental situation. All three $U(x)$ give rise to the same qualitative picture, i.e. the appearance of the non-chiral states inside the bulk band gap. Their appearance effectively reduces the size of the band gap $\Delta$. This reduction is more pronounced in device 2 with a wider junction. As our calculations will show later, the mean free path of the kink states $L_0$ decreases with decreasing $\Delta$. Similarly, $L_0$ decreases with increasing junction width $w$.

In calculations that follow, we have adopted the cosinusoidal profile shown in Supplementary Fig. 11 (a) unless otherwise noted and systematically vary the interlayer potential difference $U$, the disorder strength W, the Fermi level $E_F$ to study their effect on the conductance of the kink states.

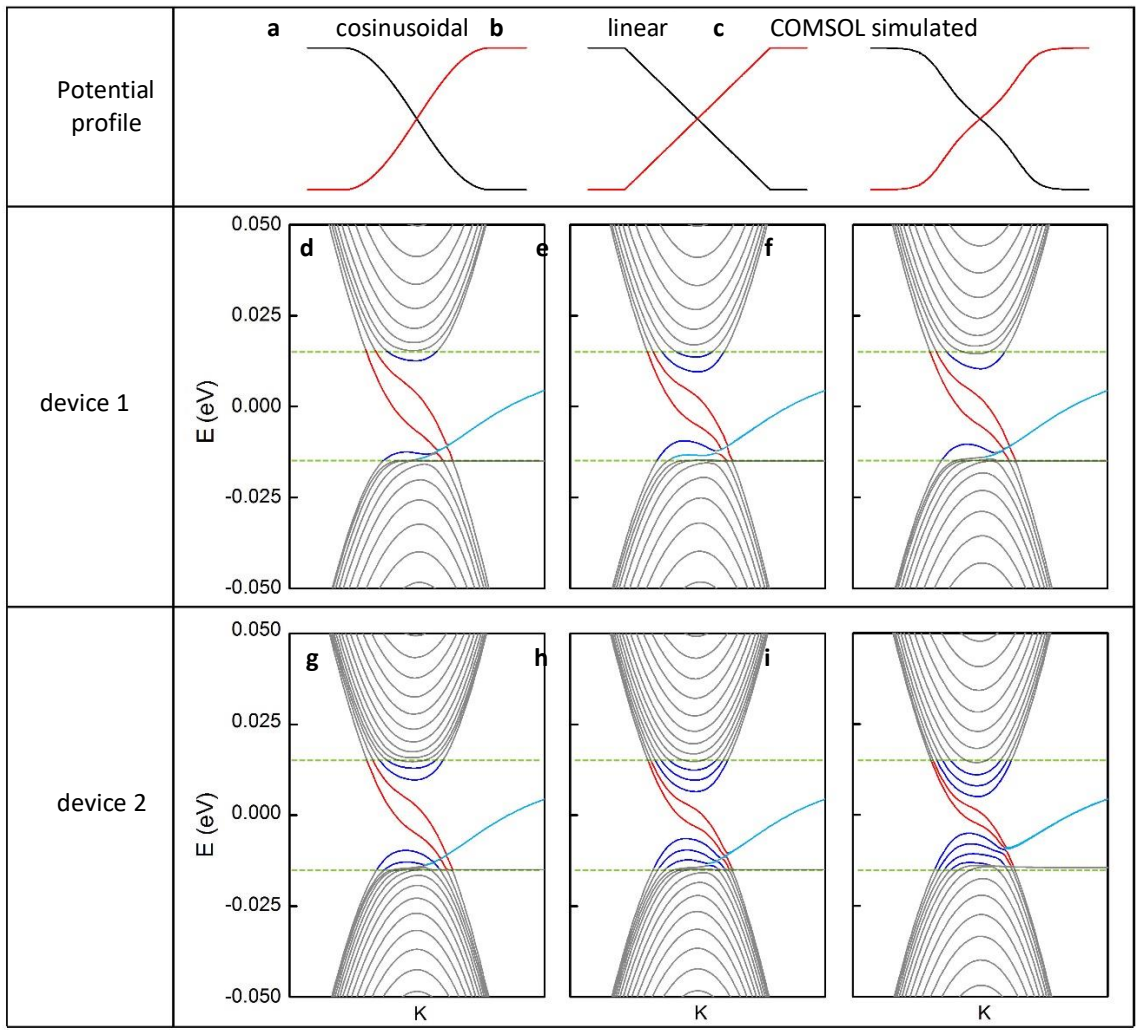

**Supplementary Figure 11** | Band structures of the bilayer graphene line junction calculated using $w = 70$ nm as in device 1 (d, e, f) and $w = 110$ nm as in device 2 (g, h, i). Three differential potential profiles are used for each width. These are respectively cosinusoidal (a), linear(b), and COMSOL simulated device potential (c, see Supplementary Fig. 2). $2U = 30$ meV in all calculations. The color scheme follows that of Supplementary Fig. 9. More bound states appear as the potential profile becomes smoother from (a) - (c). Increasing the junction width also leads to more bound states.

## 7.4. The influence of the interlayer potential difference on the band structure

Supplementary Figs. 12 (a) – 11 (c) compare three band structures calculated with different interlayer potential differences $2U$ = 30, 100 and 300 meV, respectively. Increasing $U$ leads to the increase of the bulk band gap size $\Delta$. As a result, the energy range, over which only the kink states exist, also increase. Our calculations summarized in Supplementary Table II show that the mean free path of the kink states $L_0$ increases with increasing $\Delta$, as a result of diminishing coupling to the non-chiral states. Thus, increasing the band gap can effectively suppress the backscattering of the kink states.

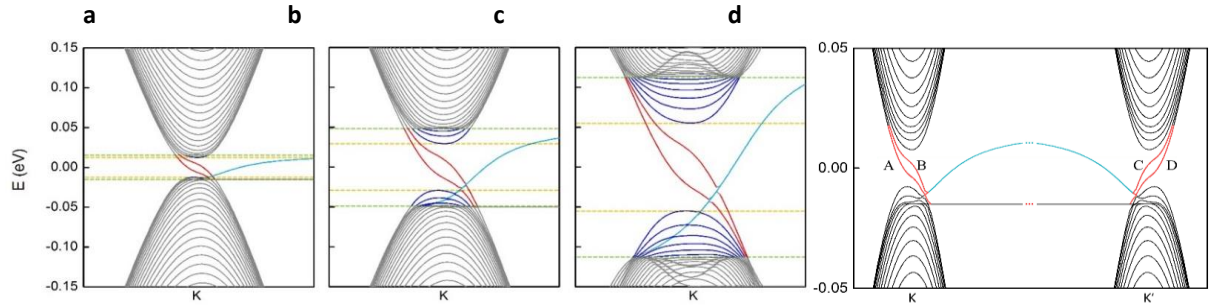

**Supplementary Figure 12** | Band structures of the bilayer graphene line junction calculated with $2U$ = 30 (a), 100 (b), and 300 meV (c). Other parameters used are the same as in Supplementary Fig. 10 (d). The color scheme follows that of Supplementary Fig. 9. Increasing $U$ increases the energy range in which only the kink states exist as marked by the yellow dashed lines. (d). The full band structure of the junction including both valleys and with the four kink modes labeled as shown. $2U$ = 30 meV.

## 8. Calculating the conductance of the kink states

### 8.1 Calculation method

With this $\pi$-band tight-binding Hamiltonian in hand, we can calculate the conductance of the bilayer graphene line junction by employing a two-terminal Landauer-Büttiker formalism based on the Green's function technique[10]:

$$G_{LR} = \frac{2e^2}{h} Tr[\Gamma_L G^r \Gamma_R G^a],$$

where $G^{r(a)}$ is the retarded (advanced) Green's function of the central scattering region. $\Gamma_{L(R)} =$

$i[\Sigma^r_{L(R)} - \Sigma^a_{L(R)}]$ is the line-width function describing the coupling between the left (right) terminal and the central scattering region, with $\Sigma^r$ being the self-energy obtained via a transfer matrix method[11]. Thirty samples are collected for each data point, the average of which is plotted.

### 8.2 The effect of disorder strength $W$

Supplementary Fig. 13 (a) plots the averaged conductance as a function of the junction length $L$ using different disorder strengths $W$ and with the same parameters used in the band structure shown in Supplementary Fig. 11 (d). We set the Fermi level $E_F = 5$ meV in the calculations. One can see that, at any fixed $W$ the conductance decreases as the junction length $L$ increases, with larger $W$ giving rise to a more rapid decrease. These results can be readily understood since on-site Anderson disorder allows inter-valley scattering events that lead to the backscattering of the kink states. Although this type of disorder is unrealistic in experimental samples, its use allows us to model potential inter-valley scattering mechanisms and examine the effect of other controlling parameters of the setup. The disorder strength $W$ is set to 0.6 eV in following calculations in order to numerically match the magnitude of the conductance observed in experiments.

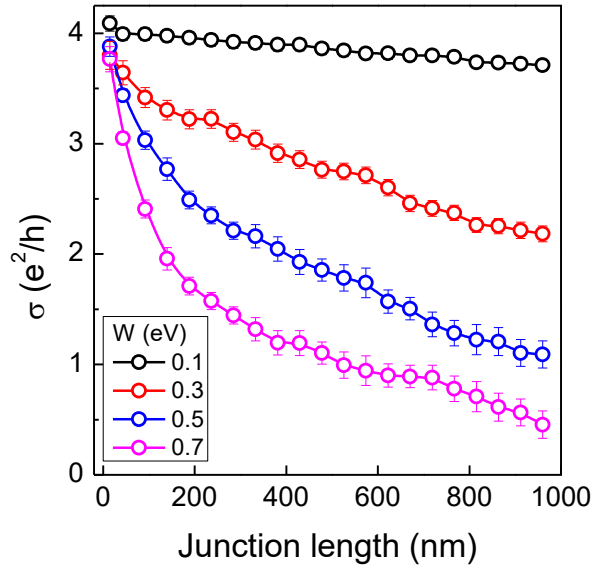

**Supplementary Figure 13** | Averaged conductance vs junction length with varying disorder strengths $W$ as indicated in the plots. See Supplementary Fig. 10 (d) for the band structure used in the calculations. $E_F$ is set to be 5 meV.

**8.3 Dependence of the conductance on the Fermi level $E_F$, the size of the band gap $\Delta$ and the junction width $w$**

The effect of the Fermi level on the conductance of the junction is studied using the parameters of device 1 (see Supplementary Fig. 11 (d), for band structure). Figure 3b of the text plots the junction conductance as a function of the junction length at different Fermi levels $E_F$ = 0, 5 and 14 meV respectively and the corresponding mean free path $L_0$. At $E_F$ = 14 meV, where one non-chiral state is also occupied, the initial junction conductance at $L$ = 0 is 8 $e^2/h$. However, it decays more rapidly with increasing $L$ compared to the other Fermi levels where only the kink states are occupied. This behavior exemplifies the important role of the non-chiral state, i.e. its co-existence enhances the backscattering probability of the kink states by providing more scattering paths that include the non-chiral states (see Fig. 3c of the text for an illustration of the effect). Thus, increasing the energy range where the kink states exist along can effectively suppress its backscattering. This can be done by increasing the size of the band gap $\Delta$ and decreasing the junction width $w$. Supplementary Table II gives the calculated $L_0$ for devices 1 at various values of $\Delta$. In addition, $L_0$ is about 25% smaller in device 2 ($w$ = 110 nm) compared to device 1 ($w$ = 70 nm). The results support the above conclusions.

**9. Modeling the effect of a perpendicular magnetic field**

When an external magnetic field $\boldsymbol{B} = -\nabla \times \boldsymbol{A}$ is applied, the tight-binding model Hamiltonian is modified by introducing a Peierls phase factor in the hopping terms[12],

$$t_{ij} \rightarrow -t_{ij} e^{-i\frac{e}{\hbar}\int \boldsymbol{A} \cdot d\boldsymbol{l}},$$

where $\int \boldsymbol{A} \cdot d\boldsymbol{l}$ is the integral of the vector potential along the path from site $j$ to $i$. In our calculations, the Landau gauge of $\boldsymbol{A} = -By\boldsymbol{e}_x$ is adopted for the perpendicular magnetic field $\boldsymbol{B} = B\boldsymbol{e}_z$.

In Supplementary Figs. 14 (a) – 13 (d), we show how the band structure shown in Supplementary Fig. 10 (d) evolves in an external magnetic field up to 12 T. As the magnetic field increases, the non-chiral states become Landau levels and lift away from the energy range of the kink states. As discussed earlier, this effect helps suppress the backscattering of the kink states.

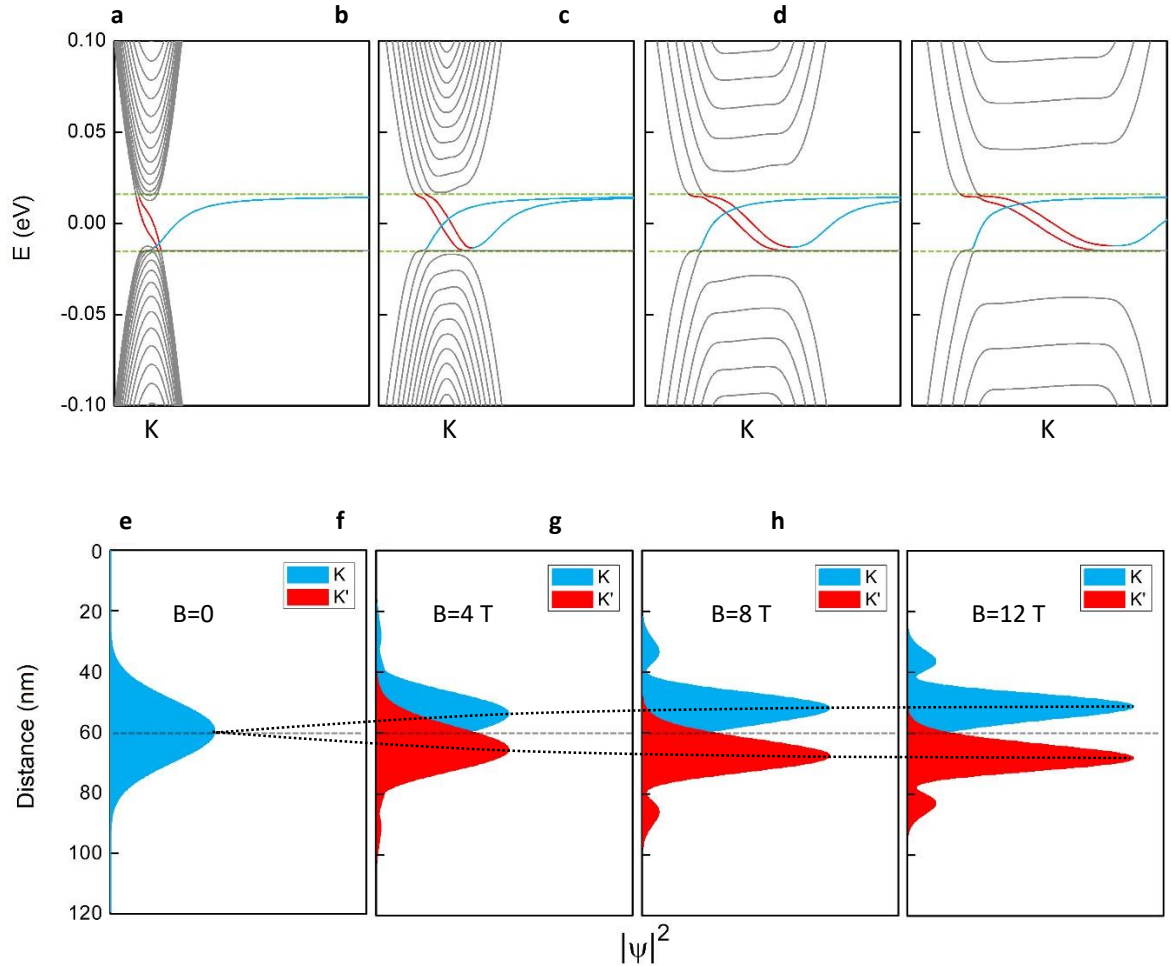

**Supplementary Figure 14** | Band structure of the bilayer graphene line junction at different magnetic fields $B = 0$ (a), 4 (b), 8 (c), and 12 T (d). See Supplementary Fig. 10 (d) for other parameters of the calculation. The formation of the Landau levels lifts the bound states away from the energy range of the kink states (red). (e) - (h) The corresponding wave function distributions of the kink states in $K$ and $K'$ valleys (modes A and D labeled in Supplementary Fig. 11(d) respectively) showing increasing separation as the magnetic field increases. $E_F = 5$ meV.

The magnetic field also introduces a second important effect that can directly reduce the coupling of the kink states in different valleys, i.e. the separation of their wave functions due to the Lorentz force. This separation increases with increasing field strength as Supplementary Figs. 14 (e) - (h) shows and also increases as the Fermi level $E_F$ departs from the CNP of the junction, as shown in Supplementary Fig. 15. This physical separation of the wave functions is reminiscent of the chiral quantum Hall edge states and offers the kink states robust protection against any direct backscattering. As we discussed in the text, the relatively short mean free paths of the kink states appear to be at odds with the cleanness of the devices. Valley coherence due to many-body effect may be at play. If this is the case, it can also be suppressed by the magnetic field induced wave function separation, allowing the kink states to recover the ballistic charge transport expected in single-particle theory. This is indeed what we observed experimentally in Fig. 4 of the text.

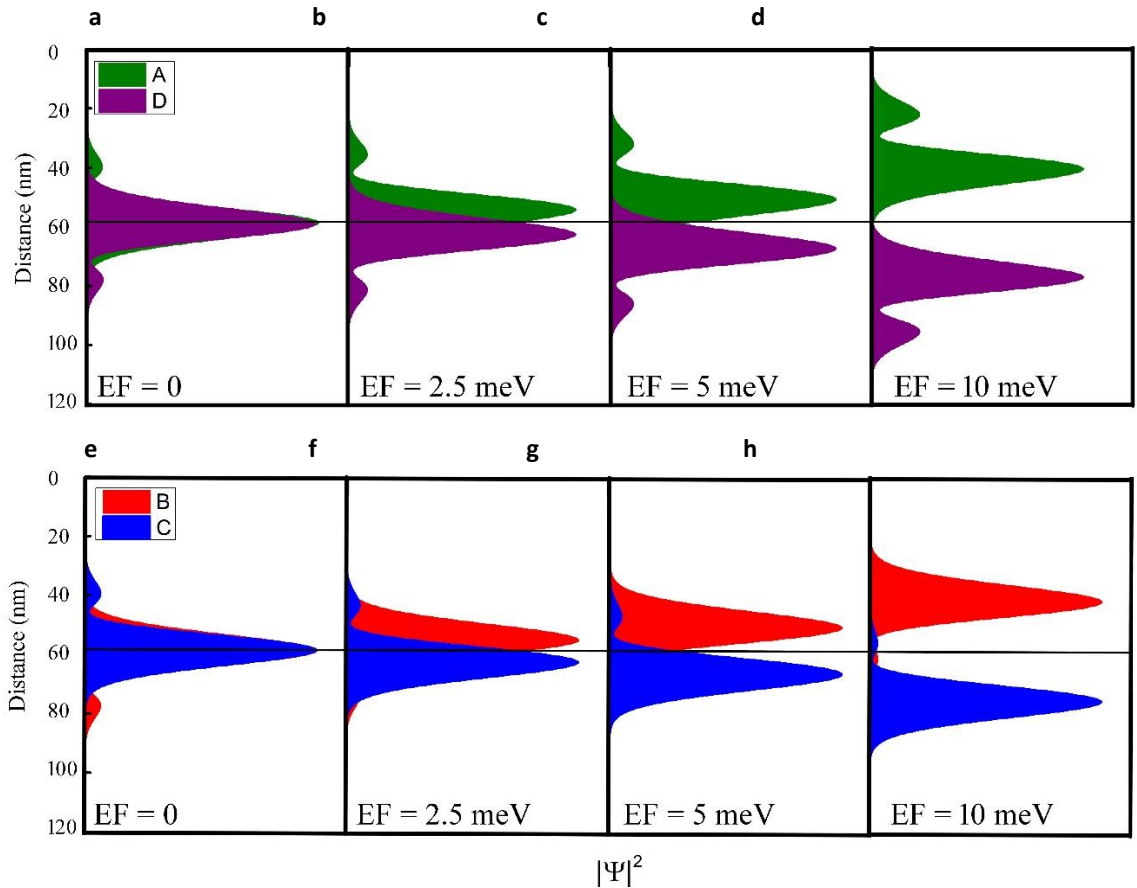

**Supplementary Figure 15 |:** Wave functions distribution of the kink states in a perpendicular magnetic field of $B = 8$ T as a function of the Fermi level $E_F$. Modes A and B are in the $K$ valley. Modes C and D are in the $K'$ valley. See Supplementary Fig. 11 (d) for the labeling scheme. Modes in the same valley largely overlap. The separation of the $K$ and $K'$ valley kink states is zero at $E_F = 0$ and increases as $E_F$ departs from the CNP of the junction.

Supplementary Fig. 16 plots the calculated junction conductance in a magnetic field. The application of a magnetic field indeed enhances the junction conductance significantly. The magneto-conductance data extracted from these curves are shown in Fig. 4d and agree reasonably well with experiment.

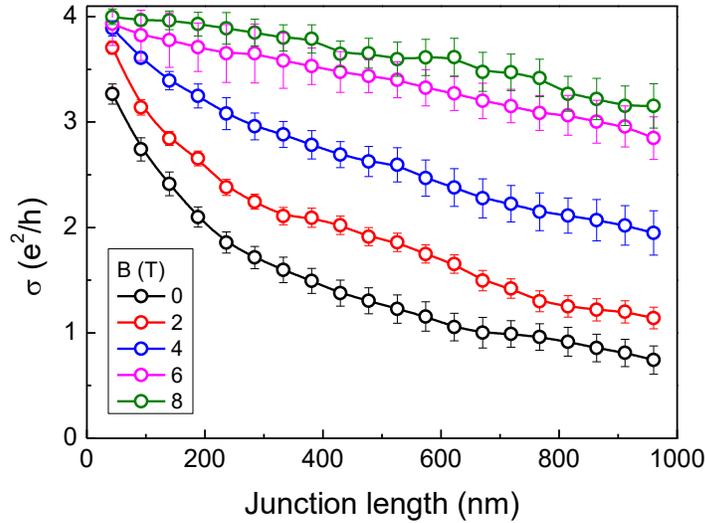

**Supplementary Figure 16** | Averaged conductance vs junction length at different magnetic fields as indicated in the plot. The parameters used to calculate the band structure are the same as those in Supplementary Fig. 10 (d) . $E_F$ = 5 meV. $W$ = 0.6 eV.

## 10. Determining the conductance of the kink states in a perpendicular magnetic field

In device 1, the junction area is connected to three working electrodes as shown in the inset of Supplementary Figure 5(a). We perform two-terminal resistance measurements $R_{39,39}$, three-terminal measurements $R_{39,310}$ and attribute the difference of the two to the resistance of electrode $R_9$. This value is used to approximate the resistance of electrode $R_3$, which has similar aspect ratio. $R_3$ is subtracted from the three-terminal data $R_{39,310}$ to obtain a four-terminal reading. As can be seen in Supplementary Figure 17, $R_9$ is only ~3 kΩ at $B$ = 0, which is a small fraction of the measured $R_j$ (several tens of kΩ). It however increases significantly with increasing $B$-field, reaching ~19 kΩ at $B$ = 7 T. Consequently, the uncertainty in the approximation of $R_3$ becomes increasingly important. We suspect that a slight increase of $R_{kink}$ in device 1 above 7 T may be due to this subtraction procedure. This behavior is not observed in device 2, where the junction resistance is measured using all four electrodes.

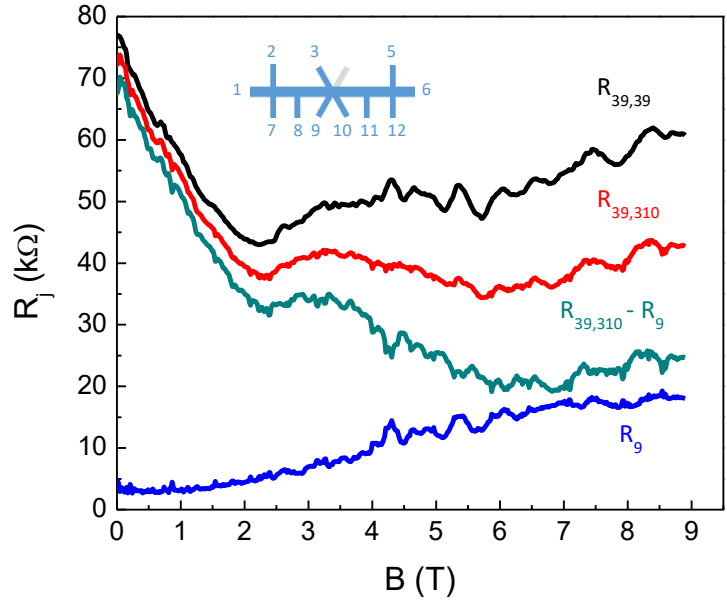

**Supplementary Figure 17 |** Magnetoresistance of the kink state in device 1. $D_L$ = +0.2 V/nm and $D_R$ = -0.3 V/nm, $V_{Si}$ = -55.5 V. The inset shows the device schematic. $R_{39,310} - R_9$ is shown in Fig. 4(a).

The resistance of the kink states in device 2 $R_{kink}$ is obtained through a two-channel model as described below. Supplementary Figures 18 (a) and (b) plot the COMSOL-simulated potential profile of the even and odd configurations. Due to the finite width of the junction ($w$ = 110 nm), the bulk gap $\Delta$ is reduced to $\Delta' \sim 1/3\ \Delta$ in the junction area in the even configuration. As Supplementary Figures 18 (c) shows, at $B = 0$, non-chiral states exist in both configurations and their band structures are nearly identical. Near $E_F = 0$, the kink states conduct in parallel with hopping conduction through charge puddles induced by Coulomb potential fluctuations of the non-chiral states, i.e. $\sigma_{j,\ total} = \sigma_{kink} + \sigma_{para}$ (Eq. S2). Here, $\sigma_{para}$ is well approximated by the junction conductance in the even configuration $\sigma_{even}$. As Fig. 4(a) shows, at $B = 0$ $\sigma_{even}$ is significant in device 2 and as a result, $\sigma_{j,\ total}$ is below $h/4e^2$ in the kink regime. As the magnetic field increases, the energies of the non-chiral states continue to move up in the odd configuration, reaching above the bulk gap $\Delta$ at 4 T. This evolution should reduce the number of puddles co-exiting with the kink states. In addition, a magnetic field suppresses hopping conduction by localizing carriers. Thus we expect $\sigma_{para}$ due to the non-chiral states to rapidly decrease with increasing $B$, making the kink state conduction increasingly dominant.

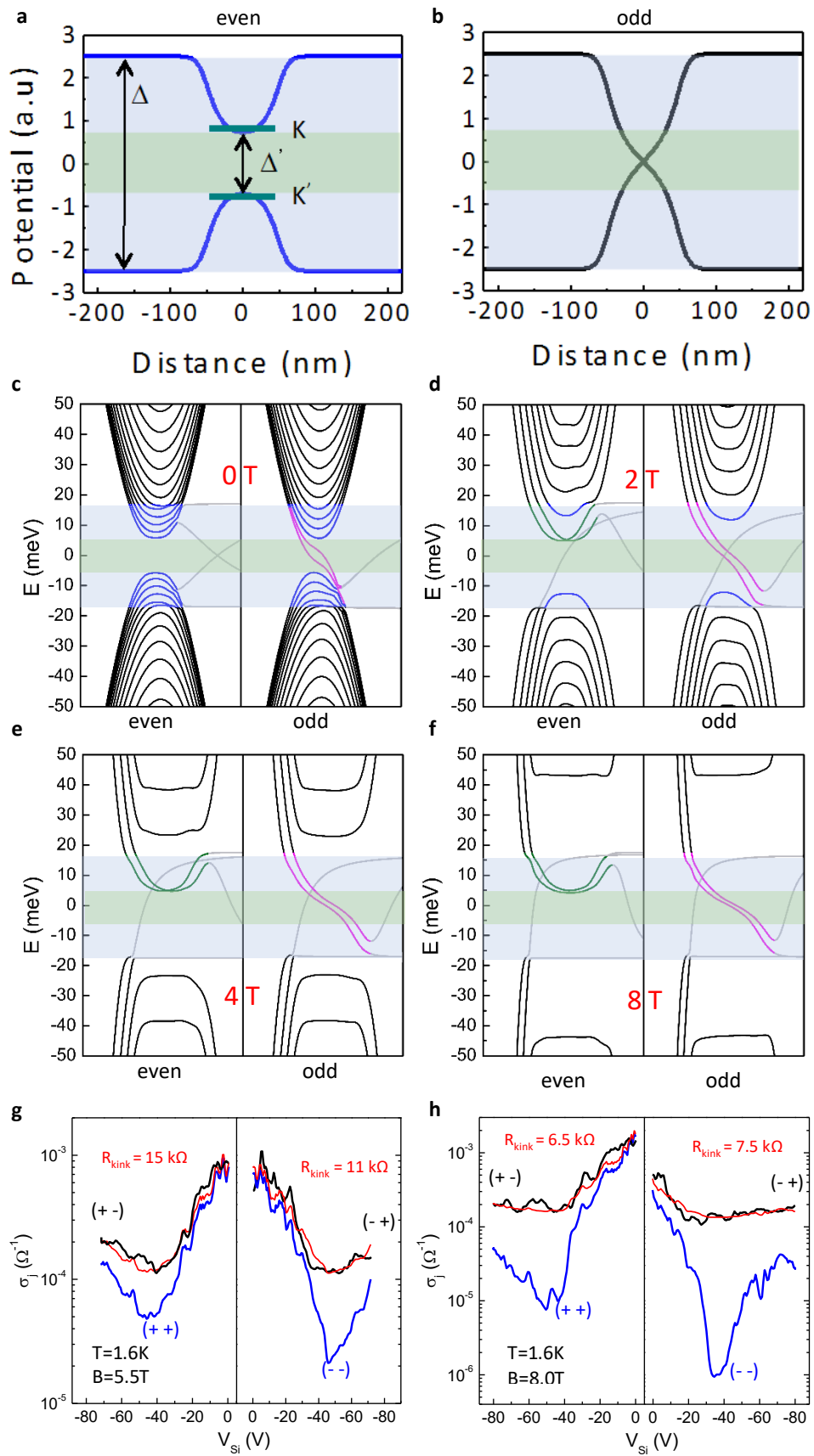

**Supplementary Figure 18** | The kink state in a magnetic field. (a) and (b) COMSOL simulated potential profile of the even (a) and odd (b) field configurations in device 2. $w$ = 110 nm. The light blue shade mark the bulk gap $\Delta$ (30 meV in (c)-(f)) and the light green shade mark the reduced gap $\Delta'$ in all figures. (c) to (f) The band structure of the junction at $B$ = 0, 2, 4 and 8 T respectively for the even and odd field configurations as labeled. Only $K$ valley is shown. The non-chiral states below the bulk gap are colored blue. The kink states magenta. The $N$ = 0 and 1 LLs of the even configurations are colored green. The quantum Hall edge states and the zigzag edge states of the system are colored gray. (g) and (h) Measured $\sigma_j$ of the even (blue) and odd (black) bias configurations. The red curves are fits of $\sigma_j$ in the odd configurations using $\sigma_{j,\text{total}} = \sigma_{\text{kink}} + \sigma_{\text{even}}$. $R_{\text{kink}} = 1 / \sigma_{\text{kink}}$ obtained from each fit is indicated in the plot.

The magnetic field dependence of the band structure in the even configuration is quite interesting. As Supplementary Figures (c)-(f) show, all non-chiral states in the even configuration behave similarly to those of the odd configuration except for one pair of states, the energies of which remain near $\pm 1/2\,\Delta'$ (The state resides at $1/2\,\Delta'$ in the $K$ valley and $-1/2\,\Delta'$ in the $K'$ valley. $K'$ valley is not shown here ). These states roughly correspond to the $N$ = 0 and 1 Landau levels of the junction area with reduced band gap $\Delta'$. Experimentally we see $R_j$ of the even configuration also rapidly increases with increasing $B$ (Inset of Fig. 4(a)). But because of these low-lying LLs, we expect the non-chiral states in the even configuration to have higher conductance than the non-chiral states in the odd configuration, i.e. $\sigma_{\text{even}} > \sigma_{\text{para}}$. Thus, approximating $\sigma_{\text{para}}$ with $\sigma_{\text{even}}$ in Eq. S2 yields a lower bound of $\sigma_{\text{kink}}$, or higher bound of $R_{\text{kink}}$. This approximation is expected to become more accurate as B approaches zero. Supplementary Figures 18 (g) and (h) give two examples of how we use Eq. S2 to obtain $\sigma_{\text{kink}}$. The two-channel model works remarkably well. We are able to fit $\sigma_{j,\text{total}}$ of the odd configuration in the entire $V_{\text{Si}}$ range using a single value of $\sigma_{\text{kink}}$ and the measured $\sigma_{\text{even}}$. The resulting $R_{\text{kink}} = 1 / \sigma_{\text{kink}}$ are plotted in Fig. 4a as open squares. They represent $R_{\text{kink}}$ accurately near $B$ = 0 and provide an upper bound of $R_{\text{kink}}$ at $B > 0$ whereas the raw data provide a lower bound of $R_{\text{kink}}$.

We note that for $B > 8$ T, both the calculated $R_{\text{kink}}$ and the raw $R_j$ merge. This is the regime where the parallel conduction of the non-chiral states becomes negligible compares to $h/4e^2$ (see inset of Fig. 4(a)) and the kink states dominate the junction conductance. In this regime, we show that $R_{\text{kink}}$ is close to $h/4e^2$ over a range of $V_{\text{Si}}$ and is independent of the bulk gap size $\Delta$ (Figure 4(b)). This observation strongly supports the nearly ballistic behavior of the kink states.

**Supplementary Table I | Device characteristics**

| Device number | | 1 | 2 |
|---|---|---|---|
| Gating efficiency ($10^{11}$ cm$^{-2}$) | TG | 10.37 | 10.24 |
| | BG | 6.77 | 9.49 |
| Unintentional doping ($10^{11}$ cm$^{-2}$) | | 0.6 | 2.4 |
| Field effect mobility (cm$^2$V$^{-1}$s$^{-1}$) | | 100,000 | 22,000 |

**Supplementary Table II | Summary of calculated mean free path**

| $\Delta$ (meV) | 220 ($w$ = 70 nm) | 100 ($w$ = 70 nm) | 30 ($w$ = 70 nm) |
|---|---|---|---|
| $E_F$ = 0 meV | - | - | 266(3) |
| $E_F$ = 5 meV | 579(21) | 394(11) | 223(4) |
| $E_F$ = 14 meV | - | - | 141(4) |